\documentclass[superscriptaddress,preprintnumbers,amssymb,nofootinbib,twocolumn]{revtex4-2}

\usepackage{tikz-feynman}
\usepackage{subfig}
\usepackage{bm}
\usepackage{esvect}
\usepackage{verbatim}
\usepackage{multirow}

\usepackage[left]{lineno}
\usepackage{blindtext}
\pdfoutput=1
\usepackage[T1]{fontenc}
\usepackage{graphicx}
\usepackage{epsfig}
\usepackage{amsmath}
\usepackage{amsfonts}
\usepackage{amssymb}
\usepackage{braket}
\usepackage{cancel}
\usepackage{pstricks}
\usepackage{color}
\usepackage{feynmf}
\usepackage[normalem]{ulem}
\usepackage{epstopdf}
\usepackage{soul}
\usepackage{cancel}

\definecolor{ao(english)}{rgb}{0.0, 0.5, 0.0}

\usepackage{braket}
\usepackage{cancel}
\usepackage{setspace}
\usepackage{pstricks}
\usepackage{xcolor}
\usepackage{float}
\usepackage{mathtools}
\usepackage{multirow}
\usepackage{booktabs}
\usepackage{enumitem}
\providecommand{\tabularnewline}{\\}

\def\missET {{\not\!\! E_T}}
\def\noeft{{\color{gray}{{\footnotesize\sf [EFT?]}}}}
\def\gsim{\lower0.5ex\hbox{$\:\buildrel >\over\sim\:$}}
\def\lsim{\lower0.5ex\hbox{$\:\buildrel <\over\sim\:$}}

\begin{document}

\title{Multi-lepton probes of new physics and lepton-universality in top-quark interactions}
\author{Yoav Afik}
\email{yoavafik@gmail.com}
\affiliation{Experimental Physics Department, CERN, 1211 Geneva, Switzerland}
\author{Shaouly Bar-Shalom}
\email{shaouly@physics.technion.ac.il}
\affiliation{Physics Department, Technion--Institute of Technology, Haifa 3200003, Israel}
\author{Kuntal Pal}
\email{kpal002@ucr.edu}
\affiliation{Physics Department, University of California, Riverside, CA 92521, USA}
\author{Amarjit Soni}
\email{adlersoni@gmail.com}
\affiliation{Physics Department, Brookhaven National Laboratory, Upton, NY 11973, USA}
\author{Jose Wudka}
\email{jose.wudka@ucr.edu}
\affiliation{Physics Department, University of California, Riverside, CA 92521, USA}

\date{\today}

\begin{abstract}
We explore the sensitivity to new physics (NP) in the associated production of top-quarks with leptons $pp \to t \bar t \ell^+ \ell^-$, which leads to the multi-leptons signals $pp \to n \ell + {\tt jets} + \not\!\! E_T$, where $n = 2,3,4$.
The NP is parameterized via 4-Fermi effective $t\bar{t} \ell^+ \ell^-$ contact interactions of various types, which are generated by multi-TeV heavy scalar, vector or tensor exchanges in $t \bar t \to \ell^+ \ell^-$; we focus on the case of $\ell=e,\mu$. 
We match the 4-Fermi $t t \ell \ell$ terms to the SMEFT operators and also give examples of specific underlying heavy physics that can generate such terms. 
Analysis of the SM signals and corresponding backgrounds shows that the di-lepton and tri-lepton channels are much better probes of the effective $t\bar{t} \ell^+ \ell^-$ 4-Fermi terms than the four-lepton one at the 13 TeV LHC.
Therefore, the best sensitivity is obtained in the di- and tri-lepton channels, for which the dominant background $pp \to t \bar t$ and $pp \to WZ$, respectively, can be essentially eliminated after applying the $2\ell$ and $ 3 \ell$ selections and a sufficiently high invariant mass selection for the opposite sign same flavor (OSSF) lepton-pair. 
We explore two cases: lepton flavor universal (LFU) NP where the $t t e e$ and $t t \mu \mu$ contact interactions are of same size and LFU violating (LFUV) NP, where the scale of the $t t \mu \mu$ terms is assumed to be much lower. 
We show that in both cases it is possible to obtain new 95\% CL bounds on the scale of the $t t \ell \ell$ contact interactions at the level 
$\Lambda \gtrsim 2-3$ TeV, which are considerably tighter than the current bounds on these 4-Fermi terms. 
\end{abstract}

\maketitle
\flushbottom

\newpage 
\tableofcontents

\section{Introduction \label{sec:intro}}

Third generation fermions are a promising window to potential new physics (NP) that underlies the Standard Model (SM). This is best manifested for the 3rd generation quark doublet and, in particular the top-quark, which is significantly heavier than all other quarks and is, therefore, expected to be the most sensitive to NP of a higher scale, such as new flavor physics~\cite{FCtopdecay1,FCtopdecay2,FCtopdecay3,FCtopdecay4,FCtopdecay5,FCtopdecay6,FCtopdecay7,FCtopdecay8,t_to_cdecay_soni,t_to_c_Hou,FCtopprod1,FCtopprod2,FCtopprod3,FCtopprod4,FCtopprod5,eetc_soni1,eetc_soni2,eetc_Hou1} and CP-violation beyond the SM~\cite{ourreview}. 
For this reason, model independent studies in the Effective Field Theory (EFT) approach have been widely applied to top physics in the past two decades \cite{Degrande:2018fog,Maltoni-global,Hartland:2019bjb,Durieux:2019rbz,singletop1,tZgamma1,SMEFTtop1,tZ3,SMEFTtop2,SMEFTtop3,SMEFTtop4,Zhang:2016omx,tH1,tH2,tH3,tH4,tH5,tH6,tH7,tH8,tZ1,tZ2,tZ3,tZ4,tZ5,tZ6,tZ7,tZ8-trilepton,tZ9-trilepton,tgamma1,tZEFT-decay1,tZprime,Gudron1,Gudron2}.
Global and comprehensive EFT studies of various types of higher dimensional operators involving the top-quark field(s) can be found in~\cite{Maltoni-global,Hartland:2019bjb,Durieux:2019rbz,SMEFTtop1,tZ3,SMEFTtop2,SMEFTtop3,Gudron2,SMEFTtop4,Ellis:2020unq,Ethier:2021bye}.
The effects of (2-quarks)(2-leptons) 4-Fermi operators (which are of interest in this study) had been recently studied also in~\cite{bsll-our,Gudron1,Gudron2,bbll-our,LFU-our,tull-our,Tonero:2020zcy,Maltoni-global,topdecay1,1008.3562,topdecay3,topdecay2,Gottardo:2019lmv}; the $tt \ell \ell$ class of operators is, however, poorly bounded as will be further discussed below.

Furthermore, persistent hints for NP involving the 3rd generation quark-doublet and lepton flavor universality violation (LFUV) have been emerging in the past decade in B-decays \cite{LHCb:2021trn,deSimone:2020kwi,Gherardi:2019zil}.
Some of the notable ones are the ratios $\Gamma(B \to K^{(*)} \mu \mu)/\Gamma(B \to K^{(*)} ee)$ ($=R_{K^{(*)}}$) and to some degree the decay $B^0_{s} \to \mu^+ \mu^-$, all associated with $b \to s \mu^+ \mu^-$ transitions, as well as the ratio $R_{D^{(*)}}$ which occurs in the SM via tree-level $b\to c \ell^- \nu_{\ell}$ (see also \cite{Aaij:2014pli,Aaij:2014ora,Aaij:2017vbb,Aaij:2015esa,Aaij:2015oid,Wehle:2016yoi,Abdesselam:2016llu,ATLAS:2017dlm,CMS:2017ivg,Bifani:2017gyn,Aaij:2019wad,Abdesselam:2019wac,Lees:2012xj,Lees:2013uzd,Huschle:2015rga,Hirose:2016wfn,Aaij:2015yra,Aaij:2017uff,Aaij:2017deq,Adamczyk:2019wyt,Abdesselam:2019dgh,Bifani:2018zmi}). 
These LFUV signals may also imply that lepton flavor violation (LFV)  
effects can be sizable, i.e., much larger than expected in the SM \cite{Glashow:2014iga}. 
In addition, the recently confirmed~\cite{gm2-recent} muon $g-2$ anomaly and
also recent interesting measurements (although with less statistical significance) that have been reported by ATLAS~\cite{ATLAS-dilepton,ATLAS-bsll} and CMS~\cite{CMS-dilepton} in 
unequal production of di-muons versus di-electrons, 
provide further hints for possible NP involved  
in high-$p_T$ lepton production, and may also  
indicate that LFUV NP may be mediated by new TeV-scale states of the underlying heavy theory; potentially in interactions between the 3rd generation quarks and the electrons and muons.

Indeed, in previous recent papers~\cite{bsll-our,bbll-our,LFU-our,tull-our} we have explored the NP effects of higher-dimensional 4-Fermi 
interactions involving 3rd generation quarks and a pair of electrons and/or muons, on scattering processes at the LHC which lead to multi-leptons final states in associations with the 3rd generation quarks. In particular, 
the flavor changing $b s \ell \ell$ leading to $pp \to \ell^+ \ell^- + j_b$ ($j_b=b$-jet)~\cite{bsll-our}, 
the SU(2) related $t c \ell \ell$ (and also $tu \ell \ell$) leading to e.g., $pp \to \ell^+ \ell^- + t$~\cite{tull-our} as well as the flavor diagonal $b b \ell \ell$ leading to e.g., $pp \to \ell^+ \ell^- + j_b$ or $pp \to \ell^+ \ell^- + 2 j_b$~\cite{bbll-our,LFU-our}.

In this paper we expand these studies and consider the effects of the (poorly constrained - see below) $tt \ell \ell$  4-Fermi contact terms on multi-lepton production in association with a top-quark pair (or a single-top) at the LHC. We note (as further discussed below) that one-loop effects of higher-dimensional 4-Fermi operators involving the top-quark, in particular the $tt \ell \ell$ ones, can also address the $g-2$ \cite{Fajfer:2021cxa,Aebischer:2021uvt} and possibly the B-physics anomalies \cite{Kamenik:2017tnu,Fox:2018ldq,Camargo-Molina:2018cwu,Ciuchini:2019usw} when the NP scale is  $O(\text{TeV})$, potentially open to direct observation. In contrast, tree-level contributions from effective operators (e.g., operators generating a $ b\to s \ell\ell$ vertex) require a NP scale in the $20-40$ TeV range. Indeed, we find that the multi-lepton signals which we study in this paper are sensitive to the $t t \ell \ell$ contact interactions if their scale is $\Lambda \sim {\rm few}$ TeV.

Finally, we emphasize that while we are, to some degree, motivated by the above mentioned few $\sigma$ deviations seen in B-physics lepton universality  tests and also in muon ($g-2$) anomaly as hints of NP, our collider based search for beyond the SM (BSM) physics in multi-lepton final states is cast in more general terms. In particular, this search is designed for both lepton flavor diagonal as well as 
off-diagonal final states,
but restricted in this study to comparing muons with electrons only. In fact such comparisons have been of interest for a very long time (see e.g.~\cite{Soni:1973pyl}).

\section{Theoretical Framework}

We adopt the parameterization used in~\cite{our_tc_paper,tull-our} and \cite{Grzadkowski:1995te,Grzadkowski:1997cj} for the effective Lagrangian of the flavor changing (FC)  $t \bar u \ell^+ \ell^- $ and flavor diagonal $ t \bar t \ell^+ \ell^-$ contact terms, respectively:
\begin{widetext}
\begin{eqnarray}
{\cal L}_{tt\ell \ell} =  {1\over\Lambda_\ell^2} \sum_{i,j=L,R} \biggl[ V_{ij}^\ell \left({\bar \ell} \gamma_\mu P_i \ell \right) \left( \bar t \gamma^\mu P_j t \right)  + S_{ij}^\ell \left( {\bar \ell} P_i \ell \right) \left( \bar t P_j t \right)  + T_{ij}^\ell \left( {\bar \ell} \sigma_{\mu \nu} P_i \ell \right) \left( \bar t \sigma_{\mu \nu} P_j t \right) \biggr] \label{4fermimatrix}~,
\end{eqnarray}
\end{widetext}
where $P_{L,R} = (1 \mp \gamma_5)/2$ 
and $V_{ij}^\ell$, $S_{ij}^\ell$, $T_{ij}^\ell$ are the dimensionless couplings of the vector, scalar and tensor 4-Fermi interactions, respectively. As mentioned before throughout this work we will focus only on the NP involving electrons and muons, i.e., the $t t \mu \mu$ and $t t e e $ terms, and assume that the scale of the corresponding LFV, off-diagonal in flavor, 4-Fermi interactions, $tt \mu e$, is much higher so that the effects of such LFV 4-Fermi operators can be neglected.

The scale of the underlying NP involving electrons and muons may be different; a lower scale for the NP involving muons provides a very reasonable interpretation of the above mentioned anomalies in B-decays and the muon $g-2$.
Thus, in what follows, we will explore two cases: 
LFU NP where $\Lambda_e = \Lambda_{\mu}$ and LFUV muon-filic NP where $\Lambda_{\mu} \ll \Lambda_e$, for which case the leading effect of the tree-level exchanges of the heavy states is manifest in the $tt \mu \mu$ contact terms.

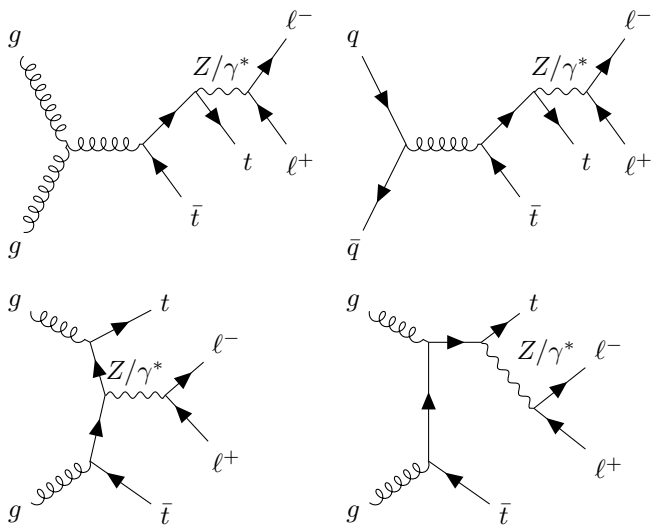
\begin{figure}[htb]
  \centering
\begin{tikzpicture}
  \begin{feynman}

    \vertex (a11) {\(g\)};
    \vertex[below=4em of a11] (b11);
    \vertex[below=8em of a11] (c11) {\(g\)};
    \vertex[right=0.7cm of b11] (b12);
    \vertex[right=1.0cm of b12] (b13);
    \vertex[right=0.7cm of b13] (b14);
    \vertex[above=2em of b14] (d14);
    \vertex[below=2em of b14] (e14) {\(\bar t\)};
    \vertex[right=0.7cm of d14] (d15);
    \vertex[below=2em of d15] (b15) {\(t\)};
    \vertex[right=0.7cm of d15] (d16);
    \vertex[above=2em of d16] (a16) {\(\ell^-\)};
    \vertex[below=2em of d16] (c16) {\(\ell^+\)};

    \vertex[right=4.5cm of a11] (a21) {\(q\)};
    \vertex[below=4em of a21] (b21);
    \vertex[below=8em of a21] (c21) {\(\bar q\)};
    \vertex[right=0.7cm of b21] (b22);
    \vertex[right=1.0cm of b22] (b23);
    \vertex[right=0.7cm of b23] (b24);
    \vertex[above=2em of b24] (d24);
    \vertex[below=2em of b24] (e24) {\(\bar t\)};
    \vertex[right=0.7cm of d24] (d25);
    \vertex[below=2em of d25] (b25) {\(t\)};
    \vertex[right=0.7cm of d25] (d26);
    \vertex[above=2em of d26] (a26) {\(\ell^-\)};
    \vertex[below=2em of d26] (c26) {\(\ell^+\)};

    \vertex[below=10em of a11] (a31) {\(g\)};
    \vertex[below=8em of a31] (d31) {\(g\)};
    \vertex[below=1.5em of a31] (b31);
    \vertex[right=1.0cm of b31] (b32);
    \vertex[above=2.0em of d31] (c31);
    \vertex[right=1.0cm of c31] (c32);
    \vertex[right=2.0cm of a31] (a33) {\(t\)};
    \vertex[right=2.0cm of d31] (d33) {\(\bar t\)};
    \vertex[below=2.0em of b32] (e32);
    \vertex[right=0.2cm of e32] (e33);
    \vertex[right=0.8cm of e33] (e34);
    \vertex[right=1.5cm of b32] (b33) {\(\ell^-\)};
    \vertex[right=1.5cm of c32] (c33) {\(\ell^+\)};

    \vertex[below=10em of a21] (a41) {\(g\)};
    \vertex[below=8em of a41] (d41) {\(g\)};
    \vertex[below=1.5em of a41] (b41);
    \vertex[right=1.0cm of b41] (b42);
    \vertex[above=2.0em of d41] (c41);
    \vertex[right=1.0cm of c41] (c42);
    \vertex[right=0.7cm of b42] (b43);
    \vertex[right=2.0cm of d41] (d43) {\(\bar t\)};
    \vertex[right=2.4cm of a41] (a44) {\(t\)};
    \vertex[below=4em of a44] (e44);
    \vertex[right=1.0cm of e44] (e45);
    \vertex[above=1.5em of e45] (e46) {\(\ell^-\)};
    \vertex[below=1.5em of e45] (e47) {\(\ell^+\)};

    \diagram* {
      (a11) -- [gluon] (b12),
      (c11) -- [gluon] (b12),
      (b12) -- [gluon] (b13),
      (b13) -- [fermion] (d14),
      (b13) -- [anti fermion] (e14),
      (d14) -- [boson, edge label=\(Z/\gamma^*\)] (d15),
      (d14) -- [fermion] (b15),
      (d15) -- [fermion] (a16),
      (d15) -- [anti fermion] (c16),

      (a21) -- [fermion] (b22),
      (c21) -- [anti fermion] (b22),
      (b22) -- [gluon] (b23),
      (b23) -- [fermion] (d24),
      (b23) -- [anti fermion] (e24),
      (d24) -- [boson, edge label=\(Z/\gamma^*\)] (d25),
      (d24) -- [fermion] (b25),
      (d25) -- [fermion] (a26),
      (d25) -- [anti fermion] (c26),

      (a31) -- [gluon] (b32),
      (d31) -- [gluon] (c32),
      (c32) -- [fermion] (e33),
      (e33) -- [fermion] (b32),
      (c32) -- [anti fermion] (d33),
      (b32) -- [fermion] (a33),
      (e33) -- [boson, edge label=\(Z/\gamma^*\)] (e34),
      (e34) -- [fermion] (b33),
      (e34) -- [anti fermion] (c33),

      (a41) -- [gluon] (b42),
      (d41) -- [gluon] (c42),
      (c42) -- [fermion] (b42),
      (b42) -- [fermion] (b43),
      (c42) -- [anti fermion] (d43),
      (b43) -- [fermion] (a44),
      (b43) -- [boson, edge label=\(Z/\gamma^*\)] (e44),
      (e44) -- [fermion] (e46),
      (e44) -- [anti fermion] (e47),

    };
    
  \end{feynman}
\end{tikzpicture}\\
\caption{Representative lowest-order SM Feynman diagrams for top-quark pair + di-lepton production, 
$pp \to t \bar t \ell^+ \ell^-$.
}
\label{fig:SM}
\end{figure}

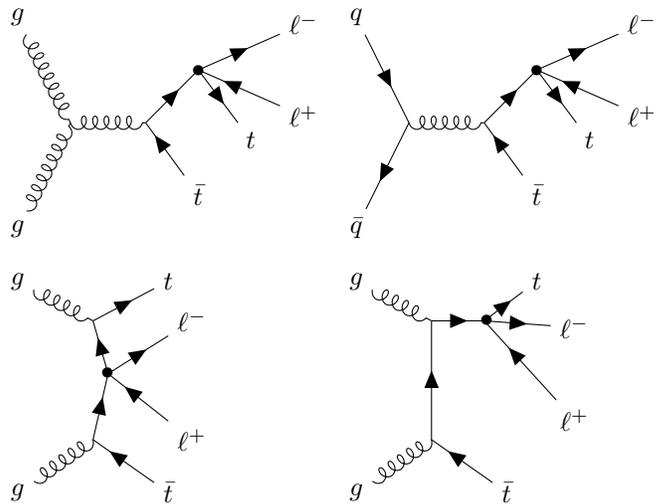
\begin{figure}[htb]
  \centering
\begin{tikzpicture}
  \begin{feynman}
    \vertex (a11) {\(g\)};
    \vertex[below=4em of a11] (b11);
    \vertex[below=8em of a11] (c11) {\(g\)};
    \vertex[right=0.7cm of b11] (b12);
    \vertex[right=1.0cm of b12] (b13);
    \vertex[right=0.7cm of b13] (b14);
    \vertex[above=2em of b14] (d14);
    \vertex[above=1.4em of b14] (d14b) {\(\bullet\)};
    \vertex[below=2em of b14] (e14) {\(\bar t\)};
    \vertex[right=0.7cm of d14] (d15);
    \vertex[below=2em of d15] (b15) {\(t\)};
    \vertex[right=0.7cm of d15] (d16);
    \vertex[above=1.0em of d16] (a16) {\(\ell^-\)};
    \vertex[below=1.0em of d16] (c16) {\(\ell^+\)};

    \vertex[right=4.5cm of a11] (a21) {\(q\)};
    \vertex[below=4em of a21] (b21);
    \vertex[below=8em of a21] (c21) {\(\bar q\)};
    \vertex[right=0.7cm of b21] (b22);
    \vertex[right=1.0cm of b22] (b23);
    \vertex[right=0.7cm of b23] (b24);
    \vertex[above=2em of b24] (d24);
    \vertex[above=1.4em of b24] (d24b) {\(\bullet\)};
    \vertex[below=2em of b24] (e24) {\(\bar t\)};
    \vertex[right=0.7cm of d24] (d25);
    \vertex[below=2em of d25] (b25) {\(t\)};
    \vertex[right=0.7cm of d25] (d26);
    \vertex[above=1.0em of d26] (a26) {\(\ell^-\)};
    \vertex[below=1.0em of d26] (c26) {\(\ell^+\)};

    \vertex[below=10em of a11] (a31) {\(g\)};
    \vertex[below=8em of a31] (d31) {\(g\)};
    \vertex[below=1.5em of a31] (b31);
    \vertex[right=1.0cm of b31] (b32);
    \vertex[above=2.0em of d31] (c31);
    \vertex[right=1.0cm of c31] (c32);
    \vertex[right=2.0cm of a31] (a33) {\(t\)};
    \vertex[right=2.0cm of d31] (d33) {\(\bar t\)};
    \vertex[below=2.0em of b32] (e32);
    \vertex[right=0.2cm of e32] (e33);
    \vertex[right=-0.02cm of e32] (e33b) {\(\bullet\)};
    \vertex[right=0.8cm of e33] (e34);
    \vertex[right=1.0cm of b32] (b33) {\(\ell^-\)};
    \vertex[right=1.0cm of c32] (c33) {\(\ell^+\)};

    \vertex[below=10em of a21] (a41) {\(g\)};
    \vertex[below=8em of a41] (d41) {\(g\)};
    \vertex[below=1.5em of a41] (b41);
    \vertex[right=1.0cm of b41] (b42);
    \vertex[above=2.0em of d41] (c41);
    \vertex[right=1.0cm of c41] (c42);
    \vertex[right=0.7cm of b42] (b43);
    \vertex[right=0.52cm of b42] (b43b) {\(\bullet\)};
    \vertex[right=2.0cm of d41] (d43) {\(\bar t\)};
    \vertex[right=2.4cm of a41] (a44) {\(t\)};
    \vertex[below=4em of a44] (e44);
    \vertex[right=0.5cm of e44] (e45);
    \vertex[above=1.5em of e45] (e46) {\(\ell^-\)};
    \vertex[below=0.5em of e45] (e47) {\(\ell^+\)};

    \diagram* {

      (a11) -- [gluon] (b12),
      (c11) -- [gluon] (b12),
      (b12) -- [gluon] (b13),
      (b13) -- [fermion] (d14),
      (b13) -- [anti fermion] (e14),
      (d14) -- [fermion] (b15),
      (d14) -- [fermion] (a16),
      (d14) -- [anti fermion] (c16),

      (a21) -- [fermion] (b22),
      (c21) -- [anti fermion] (b22),
      (b22) -- [gluon] (b23),
      (b23) -- [fermion] (d24),
      (b23) -- [anti fermion] (e24),
      (d24) -- [fermion] (b25),
      (d24) -- [fermion] (a26),
      (d24) -- [anti fermion] (c26),

      (a31) -- [gluon] (b32),
      (d31) -- [gluon] (c32),
      (c32) -- [fermion] (e33),
      (e33) -- [fermion] (b32),
      (c32) -- [anti fermion] (d33),
      (b32) -- [fermion] (a33),
      (e33) -- [fermion] (b33),
      (e33) -- [anti fermion] (c33),

      (a41) -- [gluon] (b42),
      (d41) -- [gluon] (c42),
      (c42) -- [fermion] (b42),
      (b42) -- [fermion] (b43),
      (c42) -- [anti fermion] (d43),
      (b43) -- [fermion] (a44),
      (b43) -- [fermion] (e46),
      (b43) -- [anti fermion] (e47),

    };
    
  \end{feynman}
\end{tikzpicture}\\
\caption{Representative Feynman diagrams of ${\cal O}(1/\Lambda^2)$ for top-quark pair + di-lepton production $pp \to t \bar t \ell^+ \ell^-$. The dimension six 4-Fermi interaction is marked by a heavy dot.}
  \label{fig:Feynman}
\end{figure}

We will consider below the effects of these new $tt\ell \ell$ contact interactions  ($\ell = \mu,e$) in top-quark pair production in association with a pair of opposite-sign (OS) leptons (see Figs.~\ref{fig:SM} and \ref{fig:Feynman} for representative SM and NP diagrams for these processes, respectively):\footnote{The single-top production channel in association with di-leptons: $pp \to t  \ell^+ \ell^- + j_b +j$ (and the charged conjugate channel) via the underlying $Wb$ scattering process $Wb \to t \to t \ell^+ \ell^-$, is also generated by the new $tt \ell \ell$ interactions in~Eq.~\ref{4fermimatrix} and 
are included in our analysis, although their relative contribution to the di- and tri-leptons signals considered here is considerably smaller.}
\begin{eqnarray}
&& pp \to t \bar t \ell^+ \ell^- ~, \nonumber \\
&& pp \to t(\bar t) \ell^+ \ell^- + j_b +j ~, 
\label{top-proc}
\end{eqnarray}
leading to the following di-lepton, tri-lepton or four-leptons signals ($j=$light-jet and $j_b=b$-jet):
\begin{eqnarray}
(t t \ell \ell)_{2 \ell} &\equiv& pp \to \ell^+ \ell^- + 2 \cdot j_b + 4 \cdot j ~, \nonumber \\
(t t \ell \ell)_{3 \ell} &\equiv& pp \to \ell^{\prime \pm} \ell^+ \ell^- + 2 \cdot j_b + 2 \cdot j + \missET ~, \nonumber \\
(t t \ell \ell)_{4 \ell} &\equiv& pp \to \ell^{\prime \pm} \ell^{\prime \prime \mp} \ell^+ \ell^- + 2 \cdot j_b + \missET ~, \nonumber \\
(t \ell \ell)_{2 \ell} &\equiv& pp \to \ell^+ \ell^- + 2 \cdot j_b + 3 \cdot j ~, \nonumber \\
(t \ell \ell)_{3 \ell} &\equiv& pp \to \ell^{\prime \pm} \ell^+ \ell^- + 2 \cdot j_b + j + \missET ~, 
\label{lepton_chan}
\end{eqnarray}
depending on the top-quarks decay channels: $(t t \ell \ell)_{2 \ell}$ when the $t \bar t$ pair decay hadronically via $t \bar t \to b \bar b W^+ W^- \to 2j_b +4j$, $(t t \ell \ell)_{3 \ell}$ when one top decays leptonically via $t \to bW \to j_b + \ell^\prime + \missET$ and $(t t \ell \ell)_{4 \ell}$ when both tops decay leptonically $t \bar t \to b \bar b W^+ W^- \to 2j_b + \ell^{\prime \pm} \ell^{\prime \prime \mp} + \missET$. Similarly, in the single-top channels: $(t \ell \ell)_{2 \ell}$ when the top (or anti-top) decays hadronically and $(t \ell \ell)_{3 \ell}$ when it decays leptonically. 
For example, in the LFUV muon-filic NP case we have
$2 \ell = \mu^+ \mu^-$, $3 \ell = \mu^\pm \mu^+ \mu^-, e^\pm \mu^+ \mu^-$ and $4 \ell = \mu^+ \mu^- \mu^+ \mu^-, e^+ e^- \mu^+ \mu^-, e^\pm \mu^\mp \mu^+ \mu^-$. 
We note that the di-lepton channel was extensively studied in the past few years as a potential probe of 
high-$p_T$ NP and LFUV effects at the LHC, both experimentally \cite{ATLAS-dilepton,ATLAS-bsll,CMS-dilepton} and theoretically \cite{Marzocca,Admir,Marzocca:2020ueu,topdecay2,Gherardi:2019zil,bsll-our,bbll-our,LFU-our,tull-our,Panico:2021vav,Crivellin:2021rbf,Greljo:2021kvv}. Furthermore, the tri- and four-leptons channels were recently analyzed in searches for the classic $t \bar t Z, t \bar t W$~\cite{ttV_1,ttV_2} and $t \bar t H$~\cite{ttH_1,ttH_2} top signals, as well as the interesting 4-tops ($pp \to tt \bar t \bar t$) signal~\cite{tttt_1,tttt_2}. 
They are also useful for generic NP searches~\cite{ATLAS:2021wob,Sirunyan:2020tqm} and, although they are flavor-blind, they can also be very effectively used to search for FCNC physics in the top sector~\cite{tull-our,tZprime}.

We find that (at least at the 13~TeV LHC) the four-leptons channel, $(t t \ell \ell)_{4 \ell}$, is significantly less sensitive to the NP effect generated by the $tt\ell \ell$ 4-Fermi interactions than the di- and tri-lepton channels, $(t t \ell \ell)_{2 \ell}$ and $(t t \ell \ell)_{3 \ell}$, respectively. 
Indeed, as will be shown below, 
the new 4-Fermi $tt \ell \ell$ interactions can be very efficiently isolated from the SM background as well as from other potential sources of NP in the di- and tri-leptons 
channels $(t t \ell \ell)_{2 \ell}$ and $(t t \ell \ell)_{3 \ell}$ in~Eq.~\ref{lepton_chan}, by selecting exactly $2 \ell$ and $3\ell$ charged leptons in the final state and looking at the off-Z peak behavior of the OSSF di-leptons in $p p \to t \bar t \ell^+ \ell^-$ along with extra selections on the accompanied high-$p_T$ $b$-tagged and light-jets.

\section{Matching: SMEFT and the underlying new physics} 

The dimension six 4-Fermi operators in~Eq.~\ref{4fermimatrix} can be 
matched to the so-called SM Effective Field Theory (SMEFT) framework~\cite{EFT1,EFT2,EFT3,EFT4,EFT5}, where the higher dimensional effective operators are constructed using the SM fields and their coefficients are suppressed by inverse powers of the NP scale $\Lambda$:
\begin{eqnarray}
{\cal L} = {\cal L}_{SM} + \sum_{n=5}^\infty
\frac{1}{\Lambda^{n-4}} \sum_i \alpha_i O_i^{(n)} \label{eq:EFT1}~,
\end{eqnarray}
so that $n$ is the mass dimension of the operators $O_i^{(n)}$, which equals the canonical dimension for a decoupling and weakly-coupled heavy NP and 
$\alpha_i$ are the (Wilson) coefficients which depend on the details of the underlying heavy theory (we give below an example of matching of the EFT setup to a specific underlying heavy NP scenario). 
The relevant SMEFT operators, which are related to our $tt \ell \ell$ 4-Fermi operators in Eq.~\ref{4fermimatrix}, are ($p,r,s,t$ are flavor indices):  
\begin{eqnarray}
&&{\cal O}_{lq}^{(1)}(prst) = (\bar l_p \gamma_\mu l_r)(\bar q_s \gamma^\mu q_t) ~, \nonumber \\ 
&&{\cal O}_{lq}^{(3)}(prst)=(\bar l_p \gamma_\mu \tau^I l_r)(\bar q_s \gamma^\mu \tau^I q_t)~, \nonumber \\
&&{\cal O}_{eu}(prst)=(\bar e_p \gamma_\mu e_r)(\bar u_s \gamma^\mu u_t) ~, \nonumber \\
&&{\cal O}_{lu}(prst)=(\bar l_p \gamma_\mu l_r)(\bar u_s \gamma^\mu u_t)~, \nonumber \\ 
&&{\cal O}_{qe}(prst)=(\bar e_p \gamma^\mu e_r)(\bar q_s \gamma_\mu q_t) ~, \nonumber \\ 
&&{\cal O}_{lequ}^{(1)}(prst)=(\bar l_p^j e_r) \epsilon_{jk} (\bar q_s^k u_t)~, \nonumber \\ 
&&{\cal O}_{lequ}^{(3)}(prst)=(\bar l_p^j \sigma_{\mu\nu} e_r) \epsilon_{jk} (\bar q_s^k \sigma^{\mu\nu} u_t) ~. \label{SMEFT-opts}
\end{eqnarray}

The correspondence to this parameterization to that in Eq. \ref{4fermimatrix} is
given by:\footnote{Note that no $LL$ tensor or $LL$, $LR$ and $RL$ scalar terms are generated at dimension 6; they can, however, be generated by dimension 8 operators and thus have coefficients suppressed by $\sim ( v^2 /\Lambda^4)$, where $v=246$ GeV is the Higgs vacuum expectation value.}
\begin{eqnarray}
&& 
V_{LL}= \alpha_{\ell q}^{(1)} - 
\alpha_{\ell q}^{(3)} ,~ 
V_{LR}= \alpha_{\ell u} , ~
V_{RR}= \alpha_{e u} , ~
V_{RL}= \alpha_{q e} 
~, \cr
&& 
S_{RR}= -\alpha_{\ell e q u}^{(1)} , ~
S_{LL}=S_{LR}=S_{RL}=0 
~, \cr
&& 
T_{RR}=-\alpha_{\ell e q u}^{(3)} , ~
T_{LL}=T_{LR}=T_{RL}=0 \label{vst}~.
\end{eqnarray}
\begin{table*}[htb]
\caption{Matching leptoquarks exchanges to the 4-Fermi  
operators of ${\cal L}_{tt\ell \ell}(V,S,T)$ in~Eq.~\ref{4fermimatrix} and to the SMEFT operators in~Eq.~\ref{SMEFT-opts}. See also text and Eq.~\ref{LQ-lag}. 
\label{tab:matching}}
\begin{tabular}{c|c|c|c|c|c|}
{\bf $tt\ell \ell$ coupling} & $S_1$ & $S_3$ & $R_2$ & $U_1$ 
\tabularnewline
\hline 
\hline
\
 $\alpha_{\ell q}^{(1)} $  & $|w_L|^2/4 M^2$ & $3 |z|^2/4 M^2$ & / & $-|x|^2/2 M^2$  \tabularnewline
\cline{1-5} 
\
 $\alpha_{\ell q}^{(3)} $  & $-|w_L|^2/4 M^2$ & $|z|^2/4 M^2$ & / & $-|x|^2/2 M^2$  \tabularnewline
\cline{1-5} 
\
 $V_{LL} = \alpha_{\ell q}^{(1)} - \alpha_{\ell q}^{(3)} $  & $|w_L|^2/2 M^2$ & $|z|^2/2 M^2$ & / & /  \tabularnewline
\cline{1-5} 
\
 $V_{LR} = \alpha_{\ell u}$  & / & / & $-|y_u|^2/2 M^2$ & /  \tabularnewline
\cline{1-5} 
\
$V_{RL} = \alpha_{qe}$  & / & / & $-|y_q|^2/2 M^2$ & /  \tabularnewline
\cline{1-5} 
\
$V_{RR} = \alpha_{eu}$  & $|w_R|^2/2 M^2$ & / & / & /  \tabularnewline
\cline{1-5} 
\
$S_{RR} = -\alpha_{\ell e q u}^{(1)}$  & $- w_R w_L^\star/2M^2$ & / & $-y_q y_u^\star/2M^2$ & /  \tabularnewline
\cline{1-5} 
\
 $T_{RR} = -\alpha_{\ell e q u}^{(3)}$  & $w_R w_L^\star/8M^2$ & / & $-y_q y_u^\star/8M^2$ & /  \tabularnewline
\cline{1-5} 
\hline 
\hline 
\end{tabular}
\end{table*}

The heavy physics processes that can  generate the above dimension six effective operators at tree-level consists of exchanges of heavy vectors and scalars  or their Fierz transforms. Interesting examples are 
the scalar $S_1,~S_3,~R_2$ and vector $U_1$ leptoquarks,\footnote{Note that $R_2$ is the only scalar leptoquark that does not induce proton decay.}
which transform, respectively, as $(\bar 3,1,1/3),~(\bar 3,3,1/3),~(3,2,7/6)$ and $(3,1,2/3)$ under the $SU(3) \times SU(2) \times U(1) $ SM gauge group. These leptoquarks are particularly interesting, as they can address the persistent $R_{K^{(*)}}$, $R_{D^{(*)}}$ anomalies as well as the muon $g-2$  anomaly:  both $R_{K^{(*)}}$ and $R_{D^{(*)}}$  can 
be explained by a single $U_1$ vector leptoquark~\cite{U1-1,U1-2,U1-3,U1-4,U1-5,U1-6,U1-9,U1-10,U1-11,U1-12,U1-13},
or by the scalar leptoquark pairs $S_1,S_3$ and $S_3,R_2$~\cite{S1-S3-R2-1,S1-S3-R2-2,S1-S3-R2-3,S1-S3-R2-4,S1-S3-R2-5,S1-S3-R2-6,S1-S3-R2-7}, which can also address the muon $g-2$ discrepancy~\cite{Fajfer:2021cxa,Aebischer:2021uvt}, see also~\cite{NeubertPRL,R2_1,R2_2,R2_3,R2_4,R2_5,R2_6,R2_7,R2_n,LQ-anom1}.
These leptoquarks have the following couplings to a quark-lepton pair~\cite{Dorsner:2016wpm}:\footnote{The vector leptoquark $U_1$ can have additional $d_R \gamma_\mu e_R$ and $u_R \gamma_\mu \nu_R$ couplings, which are not relevant for our $tt \ell \ell$ operators.}
\begin{eqnarray}
{\cal L}_{Y}^{S_1} &=& w_L {\bar q}^{i,C} \ell^j \epsilon_{ij} S_1 + w_R {\bar u}^C e S_1 + \text{h.c.} ~, \nonumber \\
{\cal L}_{Y}^{S_3} &=& z {\bar q}^{i,C} \ell^j (\epsilon \tau^I)_{ij} S_3^I + \text{h.c.} ~, \nonumber \\
{\cal L}_{Y}^{R_2} &=& y_q {\bar q}^i e R_2^i + y_u {\bar u} \ell^{j} \epsilon_{ij} R_2^i   + \text{h.c.} ~, \nonumber \\
{\cal L}_{Y}^{U_1} &\supset& x {\bar q} \gamma_\mu U_1^\mu \ell + \text{h.c.} 
~, \label{LQ-lag}
\end{eqnarray}
where $i,j$ are $SU(2)$ indices and flavor indices are not specified.

Tree-level exchanges of $S1,~S_3,~R_2$ and $U_1$ among the lepton-quark pairs induce (after a Fierz transformation) some of the dimension six 4-Fermi operators in Eq.~\ref{SMEFT-opts} and, therefore, the scalar, vector and tensor operators of Eq.~\ref{4fermimatrix} (see Eq.~\ref{vst}). In Table \ref{tab:matching} we give the expressions of the operator coefficients in terms of those in Eq. \ref{LQ-lag}.\footnote{A compilation of the various additional types of NP that can induce the dimension six 4-Fermi interactions in~Eq.~\ref{SMEFT-opts} can be found in~\cite{topdecay2}.} We note the following:
\begin{itemize}
\item Tree-level exchanges of the leptoquarks $S_1$ and $R_2$ (if they couple to a top-lepton pair) can generate both the scalar and tensor $tt \ell \ell$ operators, i.e., those with  $S_{RR}$ and $T_{RR}$ couplings. In addition, $S_1$ can also generate the $V_{LL}$ and $V_{RR}$ vector operators, and $R_2$  the $V_{RL}$ and $V_{LR}$ ones. 
    \item A vector leptoquark $U_1$ does not generate any of the scalar, vector and tensor 4-Fermi $tt\ell \ell$ operators in Eq. \ref{4fermimatrix}, even though it contributes to the operators ${\cal O}_{\ell q}^{(1,\,3)}$. The reason is that $V_{LL} = \alpha_{\ell q}^{(1)} - \alpha_{\ell q}^{(3)}$ (see Eq.~\ref{vst}) and $\alpha_{\ell q}^{(1)} = \alpha_{\ell q}^{(3)}$ if ${\cal O}_{\ell q}^{(1,\,3)}$ are generated by $U_1$. Note, though, that it will generate the $V_{LL}$ terms for the corresponding down-quark operators, e.g., $\left({\bar \ell} \gamma_\mu P_i \ell \right) \left( \bar b \gamma^\mu P_j b \right)$ and/or the flavor changing $\left({\bar \ell} \gamma_\mu P_i \ell \right) \left( \bar b \gamma^\mu P_j s\right)$, for which 
$V_{LL} = \alpha_{\ell q}^{(1)} + \alpha_{\ell q}^{(3)}$, see e.g.,~\cite{Gherardi:2019zil}.
\item The scalar leptoquark $S_1$ has the same quantum numbers as 
the right-handed sbottom and its couplings 
to a quark-lepton pair are, therefore, the same as in the R-parity violating (RPV) 
superpotential. Indeed, the RPV setup involving the 3rd generation quarks (the so called RPV3 of~\cite{soniRPV}) is a natural and well motivated RPV setup (see also~\cite{RPV3_1,RPV3_2}) which is also an interesting candidate for addressing the B-physics and muon $g-2$ anomalies~\cite{NeubertPRL,soniRPV,soniRPV2,soniRPV3}. Note, though, that 
in this RPV3 framework the favored resolution to the muon $g-2$ anomaly
arises from 1-loop sneutrino exchanges~\cite{soniRPV3}; 
the sneutrino does not, however, couple to 
up-quarks and therefore cannot generate our $tt \ell \ell$ 4-Fermi operators.
\end{itemize}

It is interesting to note that the tensor 4-Fermi operator $T_{RR} \left( {\bar \mu} \sigma_{\mu \nu} P_R \mu \right) \left( \bar t \sigma_{\mu \nu} P_R t \right)$ is {\it the only} $tt\mu \mu$ operator which can explain the muon $g-2$ anomaly if 
$\Lambda \sim {\cal O}(\text{few TeV})$, its contribution is~\cite{Fajfer:2021cxa,Aebischer:2021uvt}:
\begin{eqnarray}
\Delta a_\mu \sim {\cal C} \cdot T_{RR} \cdot \frac{3 }{\pi^2} \frac{m_\mu m_t}{\Lambda^2} 
\log\left(\frac{m_t^2}{\Lambda^2} \right) 
\end{eqnarray}
where ${\cal C} \sim {\cal O}(0.1)$ if the tensor operator is generated from a tree-level exchange of the scalar leptoquark $S_1$ (see Table \ref{tab:matching}). 

Finally, some of the vector $tt \mu \mu$ operators, i.e., ${\cal O}_{\ell u}$ and ${\cal O}_{eu}$, involving right-handed top-quarks and corresponding to our $V_{LR} \left({\bar \mu} \gamma_\mu P_L \ell \right) \left( \bar t \gamma^\mu P_R t \right)$ and 
$V_{RR} \left({\bar \mu} \gamma_\mu P_R \ell \right) \left( \bar t \gamma^\mu P_R t \right)$ operators, respectively, can contribute at 1-loop to the $b \to s \ell \ell$ transitions and through RGE effects in the SMEFT framework~\cite{Camargo-Molina:2018cwu,Ciuchini:2019usw}. 
This result is of particular interest because
it requires the NP scale to be $\Lambda \sim {\cal O}(\text{TeV})$ for natural couplings $V_{LR},V_{RR} \sim {\cal O}(1)$ in order to address the anomalies measured in the $R_{K^{(\star)}}$ observables~\cite{Camargo-Molina:2018cwu,Ciuchini:2019usw}.
This suggests that the scale of these vector $tt \mu \mu$ 4-Fermi operators may be within the reach of the current LHC energies, in contrast to the case where the NP effect in $b \to s \ell \ell$ is generated at tree-level by 4-Fermi $b s \ell \ell$ operators where the NP scale required to address the B-anomalies is $\Lambda \sim 40$~TeV.     

Taking in earnest the above arguments in favor of the existence of TeV-scale scalar, tensor and vector $tt \ell \ell$ 4-Fermi interactions involving right-handed top-quarks, we will focus in the rest of this work on the $S_{RR}$, $T_{RR}$ and $V_{RR}$ 
$tt \ell \ell$ operators of Eq.~\ref{4fermimatrix}. 
We note that the $V_{RR}$ operator, i.e., constructed from the SU(2) singlet fields, is the only vector 4-Fermi which does not directly contribute to b/B-physics, see  \cite{tull-our} and discussion below.

\section{Current Bounds on $tt \ell \ell$ 4-Fermi operators}

As mentioned earlier, the $tt \ell \ell$ 4-Fermi operators of~Eq.~\ref{4fermimatrix} are in general poorly bounded, primarily since they are not accessible to the "classic" signals of the top-quark: top decays and $t \bar t$ pair-production which was not accessible to LEP2 energies and at the LHC is driven by $gg$- and $q \bar q$-fusion.
An exception is for the operators involving left-handed quark isodoublets for which gauge 
invariance relates the 
$t t \ell \ell$ and  $b b \ell \ell$ 4-Fermi FC interactions.  In particular, among the operators in~Eq.~\ref{SMEFT-opts}, the $(\bar LL)(\bar LL)$ operators ${\cal O}_{l q}^{(1)}$, ${\cal O}_{l q}^{(3)}$ and the $(\bar LR)(\bar RL)$ one ${\cal O}_{qe}$, include both the $tt \ell \ell$ and $b b \ell \ell$ interactions. Thus, referring to~Eq.~\ref{4fermimatrix}, it then follows that the $V_{LL}$ and $V_{RL}$ couplings for the $t$ and $b$ quarks are related: 
$ V_{LL}(tt\ell\ell) = \alpha_{\ell q}^{(1)}(\ell \ell 33) - \alpha_{\ell q}^{(3)}(\ell \ell 33)$, $ V_{LL}(bb\ell\ell) = -\alpha_{\ell q}^{(1)}(\ell \ell 33) - \alpha_{\ell q}^{(3)}(\ell \ell 33)$ and 
$V_{RL}(tt\ell\ell) =  V_{RL}(bb\ell\ell) = \alpha_{qe}(\ell \ell 33) $, and the corresponding scales $ \Lambda $ are, therefore, the same.\footnote{Note that the correlation between operators involving the top-quark and operators involving the $b$-quark should be taken with caution, since sign differences can lead to e.g., a cancellation of effects for operators involving $b_L$ and an enhancement for those involving $t_L$ (or vice-versa).} 
Thus, gauge invariance can be used to cast limits on the $V_{LL}$ and $V_{RL}$ $tt \ell \ell$ operators from B-decays~\cite{StraubEFT} and from high-$p_T$ di-lepton searches~\cite{Marzocca}: $\Lambda \gsim 1.5 - 2$ TeV 
for the $tt ee$ and $tt \mu \mu$ 4-Fermi $V_{LL}$ and $V_{RL}$ terms in~Eq.~\ref{4fermimatrix}, for natural ${\cal O}(1)$ correspondinf Wilson coefficient, see also~\cite{tZ3,Gudron2}.   

A recent interesting search for top-quark production with
additional leptons was preformed by CMS in~\cite{Sirunyan:2020tqm}, and was used to constrain several types of dimension six operators involving the top-quark, including the 4-Fermi $tt \ell \ell$ ones. The typical sensitivity that they obtained is $\Lambda \gsim 500$ GeV for the scalar (with $S_{RR}=1$) and vector (with $V_{ij}=1$) $tt \ell \ell$ operators in~Eq.~\ref{4fermimatrix} and $\Lambda \gsim 1$ TeV for the tensor one (with $T_{RR}=1$).

Finally, we have performed a re-interpretation of the recent 
ATLAS measurements of the $3 \ell$ and $4 \ell$ signals in~\cite{ATLAS:2021wob} 
to obtain bounds on our $tt\ell \ell$ 4-Fermi terms.
In particular, we have applied the same set of selections and kinematic cuts that was used in this search to our NP signals in these multi-lepton categories, using {\sc MadGraph5\_aMC@NLO}~\cite{madgraph5} at LO parton-level as our events generator and DELPHES for detector simulation (see below for a detailed description of our simulated event samples). We then obtain the best sensitivity using their "$3 \ell$, off-Z, $\missET > 50$ GeV" selection in the $m_{3 \ell} > 600$ GeV bin, for which ATLAS obtained a 95\% CL upper limit on the NP event yield of $N_{95}(obs.) = 14$ (see Table 5 in~\cite{ATLAS:2021eyc}). Thus, for example, our expected NP signal yield for the tensor operator with $\Lambda/\sqrt{T_{RR}} =1$ TeV
(i.e., after applying exactly the ATLAS set of cuts and selections and with ${\cal L}=140$~fb$^{-1}$), is $N_{NP}(\Lambda/\sqrt{T_{RR}} =1~{\tt TeV}) =20$, 
so that, in this case also, we obtain a bound 
of $\Lambda \gsim (20/14)^{1/4} \sim 1.1$ TeV for the tensor operator with $T_{RR} \sim 1$ and 
sub-TeV level bounds for the scale of the scalar and vector operators with natural ${\cal O}(1)$ couplings. 
Note that the NP effects which we study here (from the $tt \ell \ell$ 4-Fermi operators) have not been considered in these recent CMS and ATLAS searches. Indeed, as we show below, a significantly better sensitivity to these $tt\ell \ell$ 4-Fermi terms can be obtained from a search of the $pp \to t \bar t \ell^+ \ell^-$ process with the di-,  tri-lepton and four-leptons selections and with proper jets multiplicity selections, as well as dedicated selections on the minimum of the invariant mass of the hard di-leptons from the $tt \ell \ell$ vertex.

\section{Signal and background analysis}

\subsection{Calculation setup and numerical session}

For the cross-sections of the multi-lepton processes in~Eq.~\ref{lepton_chan} we will use an $m_{\ell \ell}^{\tt min}$ cumulative cross-section, 
selecting events with $m_{\ell \ell}^{OSSF(nt)} > m_{\ell \ell}^{\tt min}$: 
\begin{eqnarray}
\sigma^{\tt cum}_{n \ell} \equiv
\int_{m_{\ell \ell}^{OSSF(nt)} \geq m_{\ell \ell}^{\tt min}} d m_{\ell \ell} \frac{d \sigma_{n \ell}}{dm_{\ell \ell}} ~, \label{cum} 
\label{CCSX}
\end{eqnarray}
where $m_{\ell \ell}^{OSSF(nt)}$ is the invariant mass of the "none-top" OSSF di-leptons of the underlying hard process, which  
are produced from the $tt \ell \ell$ vertex and not from the top-quark decays (see also below), and
$\sigma_{n \ell}$ is the cross-section 
of the $n$-leptons final state, 
e.g., $\sigma_{3 \ell}$ corresponds to the tri-lepton signal. 
In particular, $m_{\ell \ell}^{\tt min}$ 
will be selected later on to optimize the sensitivity to the NP. 

The generic form of the cumulative 
cross-section for the multi-leptons single-top and top-pair production processes, in the presence of the $tt \ell \ell$ 4-Fermi operators, is:
\begin{eqnarray}
\sigma^{\tt cum}_{n \ell}=\sigma_{n \ell}^{\tt cum,SM} + c^{\tt INT} \cdot \sigma_{n \ell}^{\tt cum,INT} + c^{\tt NP} \cdot \sigma_{n \ell}^{\tt cum,NP} ~,
\nonumber \\ \label{NP-CSX}
\end{eqnarray}
where $\sigma_{n \ell}^{\tt cum,SM}$, $\sigma_{n \ell}^{\tt cum,INT}$ and $\sigma_{n \ell}^{\tt cum,NP}$ are the cumulative SM, SM$\times$NP interference and NP$^2$ terms, respectively, and $c^{\tt INT}$, $c^{\tt NP}$ are the corresponding dimensionless NP couplings, given by:
\begin{equation}
    \begin{array}{lcc}
    & c^{\tt INT} & c^{\tt NP} \cr
    {\tt scalar} & 0 & S_{RR}^2/\Lambda_{\tt TeV}^4 ~, \cr
    {\tt tensor} & 0 &  T_{RR}^2/\Lambda_{\tt TeV}^4 ~, \cr
    {\tt vector} & V_{ij}/\Lambda_{\tt TeV}^2 & V_{ij}^2/\Lambda_{\tt TeV}^4 ~.
    \end{array} \label{NP-couplings}
\end{equation}
In particular, no interference terms are generated for the scalar and tensor operators, 
while both constructive and destructive interference is possible for the vector operators (i.e., depending on the sign of $V_{ij}$ in~Eq.~\ref{4fermimatrix}).
As mentioned above, in the following we will study the sensitivity only to the $S_{RR},V_{RR}$ and $T_{RR}$ 4-Fermi interactions; the sensitivity and reach for the other 4-Fermi vector currents, $V_{LL,},V_{RL}$ and $V_{LR}$ is similar to that of the $V_{RR}$ operator.  
We recall that the scalar leptoquarks $S_1$ and $R_2$ can generate both $S_{RR}$ and $T_{RR}$ terms, while $S_1$ can also generate the $V_{RR}$ term in Eq.~\ref{NP-couplings}. Also, as indicated in Table \ref{tab:matching}, the vector leptoquark $U_1$ cannot generate any of the NP terms in Eq.~\ref{NP-couplings}.

\subsection{Simulated Event Samples}

All event samples were generated using {\sc MadGraph5\_aMC@NLO}~\cite{madgraph5} at LO parton-level and a dedicated universal FeynRules output (UFO) model for the EFT framework was produced using {\sc FeynRules}~\cite{FRpaper}. 
The 5-flavor scheme was used for the generation of all samples, both signal and background, with the NNPDF30LO PDF set~\cite{Ball:2014uwa}.
The default {\sc MadGraph5\_aMC@NLO} LO dynamical scale was used, which is the transverse mass calculated by a $k_T$-clustering of the final-state partons.
The events were then interfaced with the {\sc Pythia 8}~\cite{Mrenna:2016sih} parton shower.
Events of different jet-multiplicities were matched using the MLM scheme~\cite{MLM} with the default {\sc MadGraph5\_aMC@NLO} parameters and all samples were processed through {\sc Delphes 3}~\cite{deFavereau:2013fsa}, which simulates the detector effects, applies simplified reconstruction algorithms and was used for the reconstruction of electrons, muons and hadronic jets.

For the leptons (electrons and muons) the reconstruction was based on transverse momentum ($p_{\mathrm{T}}$)- and pseudo-rapidity ($\eta$)-dependent efficiency parametrization and an isolation from other energy-flow objects was applied in a cone of $\Delta R=0.4$ with a minimum $p_{\mathrm{T}}$ requirement of $25$~GeV for each lepton.
Jets were reconstructed using the anti-$k_{t}$~\cite{Cacciari:2008gp} clustering algorithm with radius parameter of $R=0.4$ implemented in FastJet~\cite{Cacciari:2011ma,Cacciari:2005hq}, and were required to have transverse momentum of $p _ {\mathrm{T} } >30$~GeV and pseudo-rapidity $\left|\eta\right|<2.4$.
In cases where a selection of a $b$-jet was used, the identification of $b$-tagged jets was done by applying a $p_ {\mathrm{T}}$-dependent weight based on the jet's associated flavor, 
and the MV2c20 tagging algorithm~\cite{ATL-PHYS-PUB-2015-022} in the 70\% working point.

The dominant types of background processes were considered, depending on the number of leptons in the final state as well as the irreducible SM background from $pp \to t \bar t Z / \gamma^*$ for all channels. In particular, we found that the dominant background for the $(tt \ell \ell)_{2\ell}$ sample is 
$pp \to t \bar t$ (where both top-quarks decay hadronically), for the $(tt \ell \ell)_{3\ell}$ sample it is $pp \to WZ$ (where both vector-bosons decay leptonically) and for the $(tt \ell \ell)_{4\ell}$ sample
it is $pp \to ZZ$ followed by $ZZ \to 4 \ell$ (see e.g.,~\cite{ATLAS:2021wob,Sirunyan:2020tqm}). 
All other SM backgrounds (e.g., $Z$+jets production in the case of the di-lepton signal and $t \bar t$ production in the tri-lepton channel as well as $t \bar t W$ for both the di- and tri-leptons channels) were found to be negligible given the selections of the analysis, as described next.\footnote{We note that the background from the $ttW$ channel is comparable to the SM irreducible background from $t \bar t Z$ in the case of the $3\ell$ and $2 \ell$ signals. However, both the $ttW$ and the $t \bar t Z$ backgrounds are much smaller than the leading $WZ$ and $t \bar t$ ones for these channels.}

\subsection{Event selection and background}

\begin{figure*}[]
\centering
\includegraphics[width=0.45\textwidth]{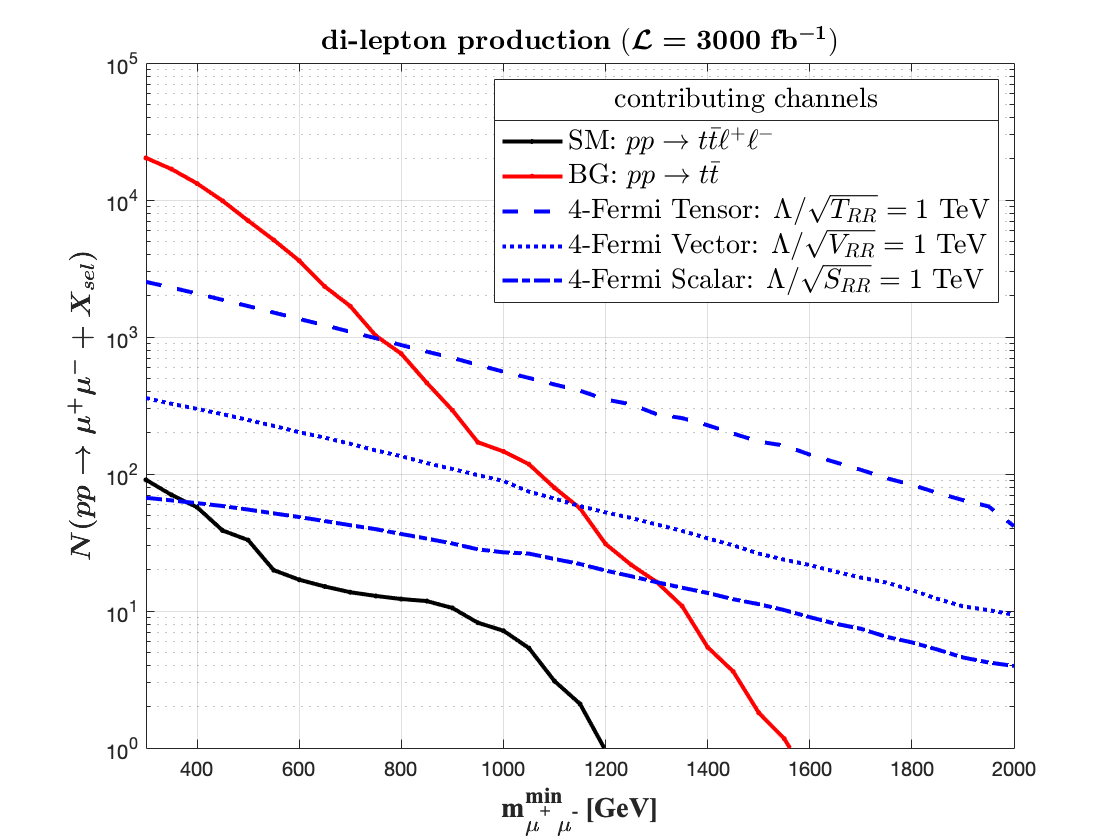}
\includegraphics[width=0.45\textwidth]{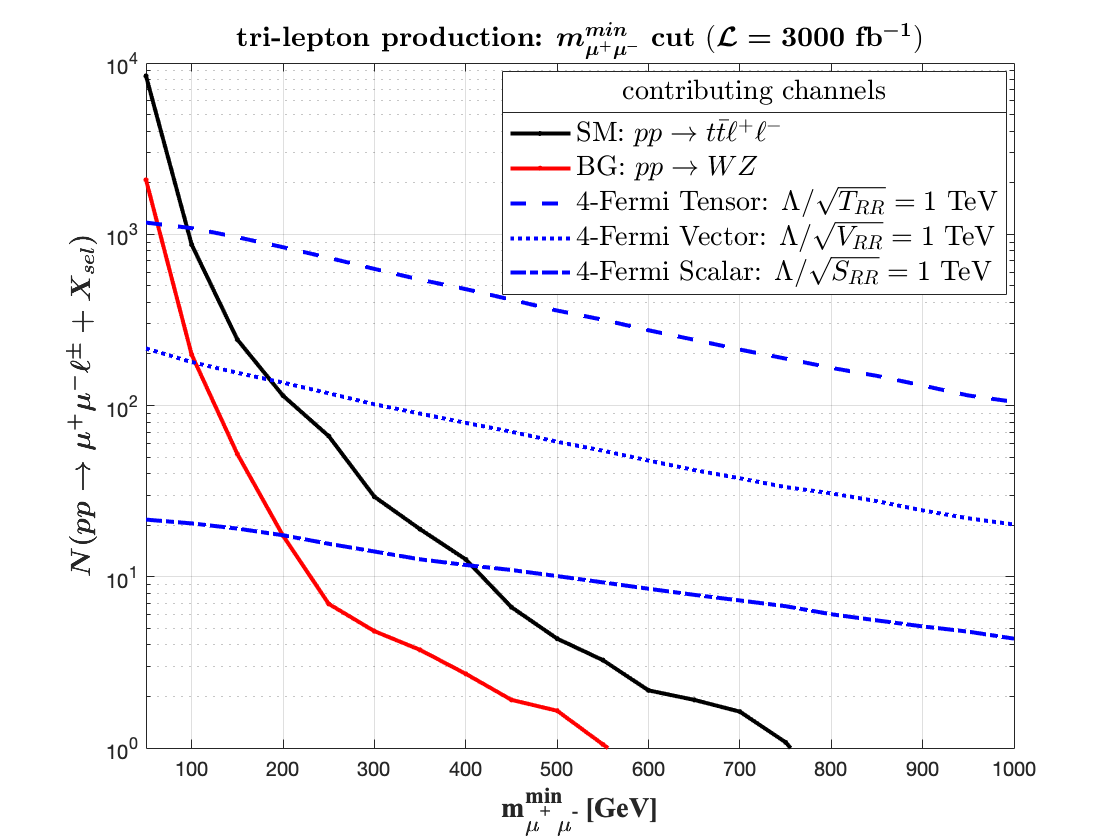}
\includegraphics[width=0.45\textwidth]{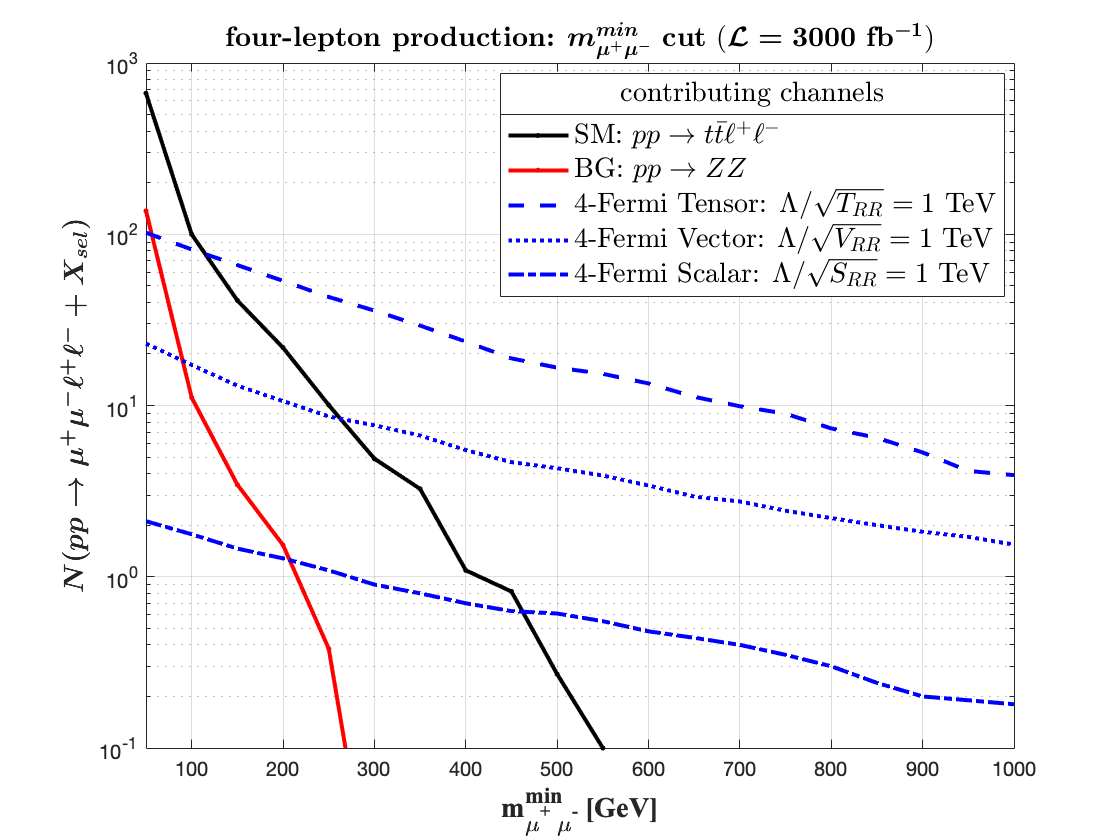}\\
\includegraphics[width=0.45\textwidth]{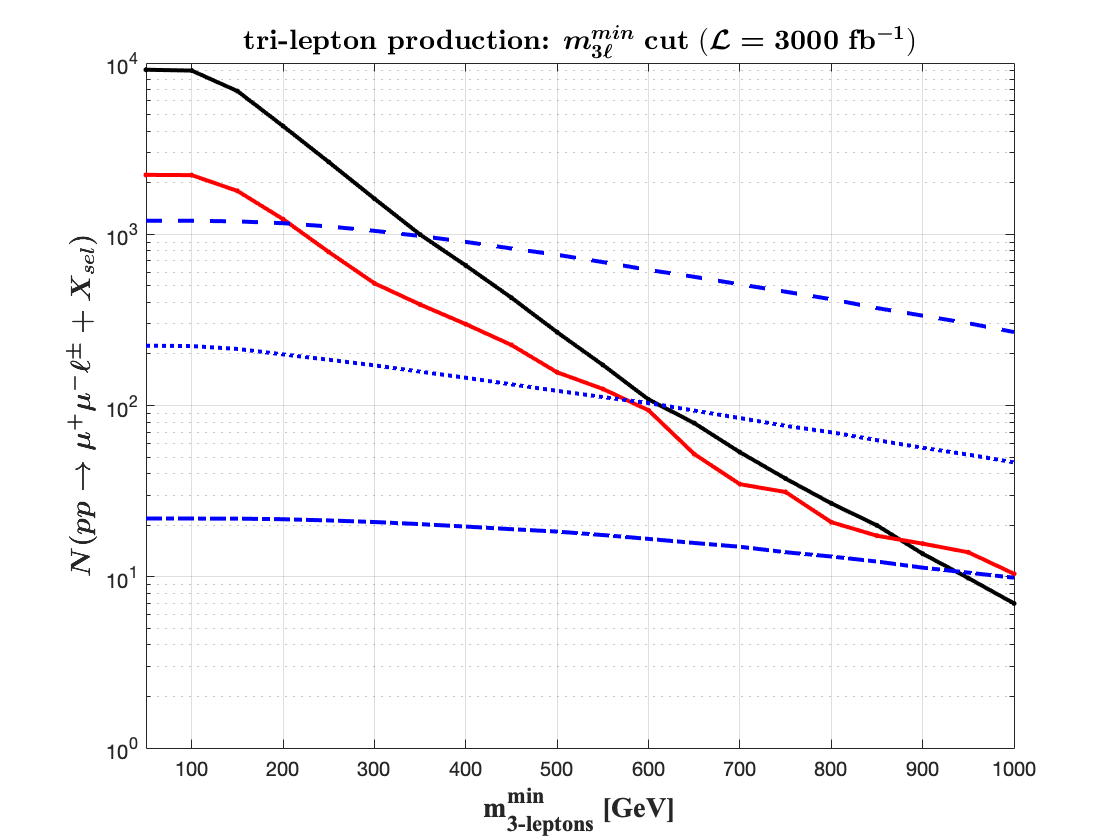}
\includegraphics[width=0.45\textwidth]{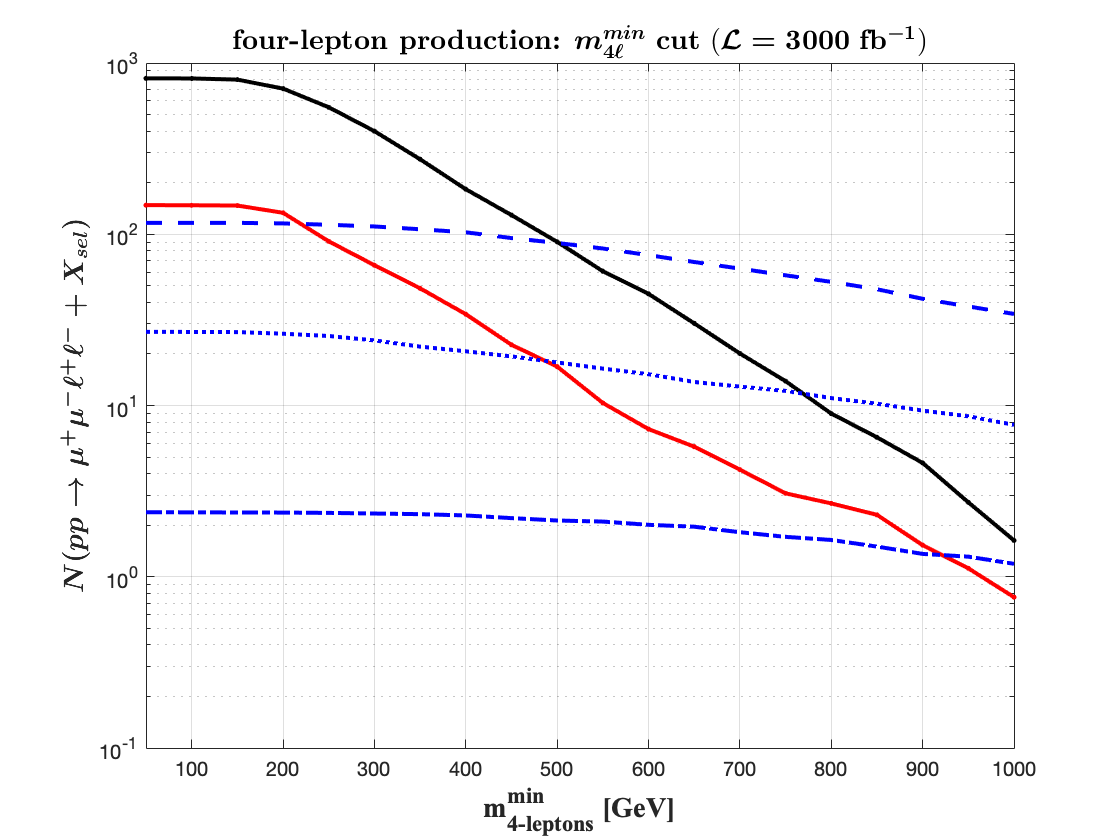}
\caption{Expected number of SM, NP and Background event yields for the di-, tri- and four-leptons signals at the HL-LHC with an integrated luminosity of ${\cal L}=3000$~fb$^{-1}$. The event yields are plotted as a function of the lower cut on the invariant mass of the none-top di-muon pair
$m_{2\ell}^{\tt min}$ (upper plots) and as a function of the lower cut on the invariant mass of all leptons in the final state in the tri-lepton and four-leptons channels  (lower plot), $m_{3\ell}^{\tt min}$ and $m_{4\ell}^{\tt min}$, respectively. The jet selections of Table \ref{tab:selections} for each channel are applied. See also text.}
\centering
\label{fig:2levents}
\end{figure*}

As noted before, in order to optimize the sensitivity to our $ttll$ 4-Fermi operators, we will isolate the NP signals in the $2 \ell$, $3 \ell$ and $4 \ell$ categories using a set of selections which are summarized in Table \ref{tab:selections}, for both the SM background and the NP signals. 
In particular, an important discriminating variable is the invariant mass of the "none-top" OSSF muon pair $m_{\ell \ell}^{OSSF(nt)}$, where the "none-top" leptons are selected to be the ones with the smallest angular separation between them, $\Delta R$. As mentioned above, we then use a minimum value of $m_{\ell \ell}^{OSSF(nt)}$, noted as $m_{\ell \ell}^{\tt min}$, and select events only if $m_{\ell \ell}^{OSSF(nt)} \geq m_{\ell \ell}^{\tt min}$, see also Eq.~\ref{CCSX}.
For the tri-lepton and four-lepton channels, we also tested  the invariant mass of all three and four leptons, respectively, as the discriminating variable, but found a reduced sensitivity to the NP in these cases, as can be seen from the lower plots  of Fig.~\ref{fig:2levents}.  This is because the leptons from the top-quark decay are considerably softer than the leptons from the "hard-process" (i.e., from the $tt \mu \mu$ vertex), so that the invariant mass spectrum of all three or four-leptons is significantly milder and is, therefore, not as effective as the none-top di-muons for isolating the NP signal from the SM background.

Finally, as indicated in Table~\ref{tab:selections}, we require a minimum number of jets and $b$-tagged jets.
For the $3\ell$ and $4\ell$ channels, we require at least two jets with at least one $b$-tagged jet; requiring a higher value of $b$-tagged jets in these samples was found to reduce the sensitivity due to selection efficiency effects.
For the $2\ell$ channel, we select at least three jets, with at least one $b$-tagged jet. Here, the selection of at least three jets was found to significantly reduce the di-leptonic $t \bar t$ background, while keeping a large fraction of the signal, as it involves a fully-hadronic top-pair decay.
\begin{table}[htb]
\caption{Selections for the tri- and four-lepton events. $m_{\ell \ell}^{OSSF(nt)}$ refers to the invariant mass of the "none-top" OSSF muon pair, which are selected to be the OSSF di-muons with the smallest angular separation between them, $\Delta R$. See also text. \label{tab:selections}}
\begin{tabular}{cccc}
 Selection & $2 \ell$ & $3 \ell$ & $4 \ell$  
\tabularnewline
\hline 
\
Number of Leptons: & exactly 2 & exactly 3  & exactly 4   \tabularnewline
Jet multiplicity: &  $\geq 3$ & $\geq 2$ & $\geq 2$  \tabularnewline
Number of b-jets: & $\geq 1$ & $\geq 1$ & $\geq 1$  \tabularnewline
$m_{\ell \ell}^{OSSF(nt)}$: & \multicolumn{2}{c}{$> m_{\ell \ell}^{\tt min}$} \tabularnewline
\hline 
\end{tabular}
\end{table}

In Fig.~\ref{fig:2levents} we plot the resulting di-, tri- and four-leptons event yields for the SM irreducible contribution and dominant backgrounds ($t \bar t$, $WZ$ and $ZZ$, respectively), as well as for the NP contributions, i.e., from the interference and pure NP parts of the cross-sections (see Eqs.~\ref{NP-CSX} and \ref{NP-couplings}), after applying the jet selections of Table \ref{tab:selections} and for an integrated luminosity of 3000~fb$^{-1}$. The event yields are plotted for the cumulative cross-section of Eq.~\ref{cum}, as a function of the lower cut on invariant mass of the none-top di-muons, $m_{\mu \mu}^{\tt min}$. To demonstrate the effectiveness of the lower cut selection on the invariant mass of the none-top di-muons, $m_{\mu \mu}^{\tt min}$, we also show in Fig.~\ref{fig:2levents} the event yields in the tri- and four-leptons channels, as a function of the lower cut on the invariant mass of {\it all} 3-leptons and 4-leptons, respectively. 

We see that in all multi-lepton channels the SM irreducible part   
and background sharply drop with $m_{\mu \mu}^{\tt min}$. In particular, we find that the SM cross-section is very sensitive to the $m_{\mu \mu}^{\tt min}$ selection, while the sensitivity of the various backgrounds (from $pp \to t \bar t,WZ,ZZ$) to the $m_{\mu \mu}^{\tt min}$ selection arises mainly due to our additional jet selections of Table \ref{tab:selections}. 

\subsection{Domain of validity of the EFT setup \label{EFTvalidity}}

The basic assumption underlying the EFT approach is that the mass of the lightest heavy particle from the underlying heavy theory is larger than $\Lambda$, 
so that none of these particles can be directly produced in the processes being investigated.  
This leads to the requirement $ \Lambda^2 \gtrsim \hat s$, where $\sqrt{\hat s}$ is the center-of-mass energy of the hard process. Alternatively, it is required that the NP cross-sections do not violate tree-level unitarity bounds, which  leads to similar constraints (for the case at hand the $tt \ell \ell$ 4-Fermi operators generate a cross-section that grows with energy $\sigma^{\tt NP}_{t \bar t \ell \ell} \propto \hat s$). These criteria, however, are not precise enough for our purposes for the following reasons:
\begin{itemize}
    \item The 4-fermion operators that we consider can be generated either by a $Z$-like heavy particle coupling to lepton and quark pairs ({\it eg.} $t t \to X \to \ell \ell$), or by a leptoquark coupling to quark-lepton pairs ({\it eg.} $t \ell \to {\tt LQ} \to t \ell$). In the first case the EFT is applicable when $\Lambda > m_{\ell\ell}^{\tt max}$ and in the second case when $\Lambda > m_{t\ell}^{\tt max}$, where $m_{t\ell}$ is the invariant mass of the top-lepton pair from the $tt \ell \ell$ contact interaction.
    \item The constraints we derive will be on the effective scale $\Lambda_{\tt eff} = \Lambda/\sqrt{f}$, whence the EFT applicability conditions become 
    $\Lambda_{\tt eff} > m_{ \ell\ell}^{\tt max}/\sqrt{f}$ or $\Lambda_{\tt eff} >  m_{q\ell}^{\tt max}/\sqrt{f}$. Thus, the EFT approach remains applicable even for situations where $\Lambda_{\tt eff}$ is of the same order, or even somewhat smaller than $m_{q\ell,\,\ell\ell}^{\tt max}$. This corresponds to NP scenarios with $ f>1$ (while still remaining perturbative). Note for example that applying the naive EFT validity criteria, $s < \Lambda_{\tt eff}^2$, to the Fermi theory of weak interactions would give $s < (246~{\rm GeV})^2$ if $f=1$, but in reality $f \sim 0.3 $ and therefore $s \lsim (100~{\rm GeV})^2$.
\end{itemize}
Based on this we can define the region of applicability by demanding $ \Lambda > m_{\ell\ell}^{\tt max}$ {\em or} $ \Lambda > m_{q\ell}^{\tt max}$, and allow $ \Lambda_{\tt eff} $ to be smaller than $m_{q\ell,\, \ell\ell}^{\tt max}$ by an $O(1)$ factor.

To close this section we note that dimension 8 operators that interfere with the SM also generate ${\cal O}(\Lambda^{-4})$ contributions to the $pp \to t \bar t \ell^+ \ell^-$ cross section. 
These, however, can be ignored compared to the ${\cal O}(\Lambda^{-4})$ NP(dim.6)$\times$NP(dim.6) terms that we keep, because the SM amplitude is much suppressed for the high $m_{\ell \ell}^{\tt min}$ selections that we use (see below).

\section{Results}
\subsection{Cut-and-Count Study}

The sensitivity to the $t t \ell \ell$ 4-Fermi NP operators is estimated using a 
"cut and count" analysis, as described below, for the di-lepton, tri-lepton and four-lepton signals. The methods used here are similar to the ones used in our previous works~\cite{bbll-our,tull-our}. 
For definiteness we will assume here that the NP couples only to muons, i.e. that the effect of the $t t \mu \mu$ operator dominates, but it should be kept in mind that a similar analysis for NP which couples to electrons will only differ in selection efficiency and detector acceptance effects. If the NP is LFU and couples equally to both electrons and muons, then we expect a slightly improved sensitivity when applying our analysis 
to the corresponding final states containing both electrons and muons.

\begin{figure}[H]
\centering
\includegraphics[width=0.52\textwidth]{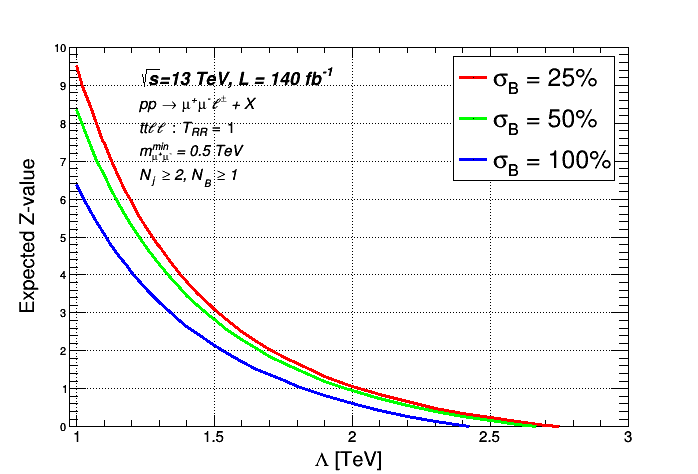}
\caption{Expected $Z$-value (defined as the number of standard deviations from the background-only hypothesis given a signal yield and background uncertainty) for the signal hypotheses varied with respect to the scale $\Lambda$ of the $tt \mu \mu$ tensor operators with $T_{RR}=1$,
for a selection of three leptons with $m_{\mu^+\mu^-}^{\tt min} = 0.5$~TeV and an integrated luminosity of 140~fb$^{-1}$.}
\centering
\label{fig:Z_value_3l_140}
\end{figure}

We calculate the expected $Z$-value, which is defined as the number of standard deviations from the background-only hypothesis given a signal yield and background uncertainty, using the \verb|BinomialExpZ| function by \verb|RooFit|~\cite{Verkerke:2003ir}.
We then find an optimized selection $m_{\mu\mu}^{\tt min}$ by maximizing the expected $Z$-value for each signal hypothesis, where at least one expected event was demanded for the signal.
An example of the expected $Z$-value from the tri-lepton signal is plotted in Fig.~\ref{fig:Z_value_3l_140}, as a function of $\Lambda$ for the case of the tensor  $t t \mu \mu$ operator ($T_{RR}=1$) and for several values of the relative overall background uncertainty, $\sigma_B=25\%,50\%$ and $100\%$, with the currently available integrated luminosity of 140~fb$^{-1}$. Clearly, the sensitivity to the NP depends on the relative uncertainty. Keeping that in mind, we analyze below all signal channels with a benchmark value of $\sigma_B = 25\%$ (see e.g., \cite{bbll-our,tull-our}), assuming that the signal uncertainty is included within $\sigma_B$.

In order to set the expected bound on the scale of NP, we calculated the $p$-value for each signal and background hypothesis using the
\verb|BinomialExpP| function by \verb|RooFit|~\cite{Verkerke:2003ir}.
In particular, we calculate the $p$-value of the background-only and background+signal hypotheses for each point and then perform a $CL_{s}$ test ~\cite{Read:2002hq} to determine the 95\% Confidence Level (CL) exclusion values for $\Lambda$.

\begin{figure}[]
\centering
\includegraphics[width=0.52\textwidth]{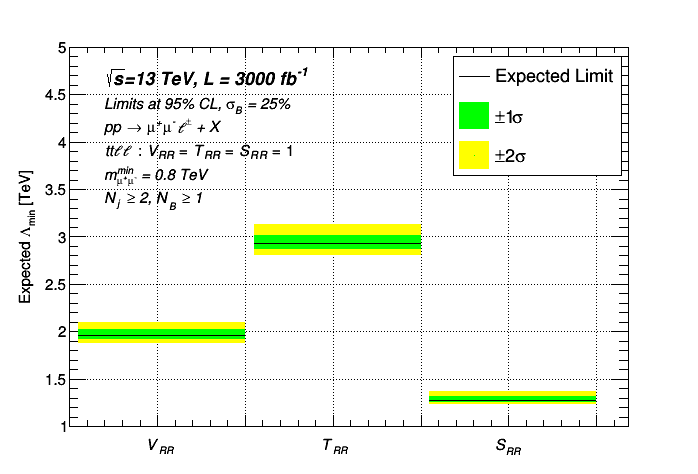}
\caption{
Expected 95\%~CL upper limit on $\Lambda$, $\Lambda_{\tt min}(95\%~CL)$, of the $tt \mu \mu$ operator for 3 signal scenarios: $S_{RR}=1$, $T_{RR}=1$ and $V_{RR}=1$, with a selection of three leptons, what a total integrated luminosity of 3000~fb$^{-1}$ and with the optimized $m_{\mu^+ \mu^-}^{\tt min} > 0.8$ TeV selection (see also Table \ref{tab:selections:ttll}).   
For all cases the overall uncertainty is chosen to be 25\% at $1\sigma$, as explained in the text.
}
\centering
\label{fig:Exclusion_3l}
\end{figure}

In Table~\ref{tab:selections:ttll} we summarize our results for the expected
95\%~CL bounds on the scale $\Lambda$ of the scalar, vector and tensor $tt \mu \mu$ operators, with natural couplings of: $S_{RR} = 1$, $V_{RR} = \pm1$ and $ T_{RR} = 1$, respectively, and for three integrated luminosity scenarios of ${\cal L}= 140$, $300$ and $3000$~fb$^{-1}$, which correspond to the data collected so far, the data expected to be available at the end of Run-3, and the data expected to be recorded at the HL-LHC. We also depict in Fig.~\ref{fig:Exclusion_3l} the 
95\%~CL bounds on the scale of the 
scalar, vector and tensor $tt \mu\mu$ operators, obtained via the tri-lepton channel at the HL-LHC with ${\cal L}= 3000$~fb$^{-1}$,
for the optimized $m_{\mu^+ \mu^-}^{\tt min}$ selection which yields the best expected limit for this case, along with the 
$\pm 1 \sigma$  and $\pm 2 \sigma$ band. 

Evidently, the optimized (best) $m_{\mu^+ \mu^-}^{\tt min}$ selection is considerably milder in the tri- and four-lepton channels, since in these cases the signal rates are reduced (partly due to the smaller branching ratio for the top to decay to leptons).
We also see that the sensitivity to the NP in the di- and tri-lepton channels is comparable with a slight advantage for the tri-lepton signal in the cases of the tensor and vector operators, where it is possible to reach a sensitivity of up to $\Lambda \sim 2(3)$ TeV with 
${\cal L} =300(3000)$~fb$^{-1}$ for the tensor case and $\Lambda \sim 1.3(2)$ TeV with 
${\cal L} =300(3000)$~fb$^{-1}$ for the vector case. Finally, notice the rather negligible sensitivity to   interference term in the vector 4-Fermi case ($\sigma_{n \ell}^{\tt cum,INT}$ in Eq.~\ref{NP-CSX}), which is due to our set of selections that are designed to minimize the SM contribution/amplitude.

Finally, we note that, in the di-lepton channel, the optimized $m_{\mu^+ \mu^-}^{\tt min}$ values exceed $\Lambda_{\tt min}/\sqrt{f}$ for $f \sim 1$, where $f=V_{RR},S_{RR},T_{RR}$ (see Table~\ref{tab:selections:ttll}). While 
this might still be within the validity regime of the EFT setup, as explained in section \ref{EFTvalidity} above, the results obtained in this channel may be "questionable" in that respect. Therefore, the better sensitivities that we obtain in the $3 \ell$ channel are also more reliable, since in this case (and also in the $4 \ell$ case) $m_{\mu^+ \mu^-}^{\tt min} < \Lambda_{\tt min}$.

In the next section we will perform a sensitivity analysis which combines the information from all three channels, i.e., the di-, tri- and four-leptons channels and which focuses on LFUV signals of the $tt \mu \mu$ 4-Fermi operators.

\begin{table*}[]
  \caption{Expected maximum 95\%~CL sensitivity ranges to the scale $\Lambda$ (denoted as $\Lambda_{\tt min}(95\%~CL)$), of the $t \bar t \mu \mu$ 4-Fermi operator, obtained via the di-, tri- and four-leptons channels with the corresponding optimal $m_{\mu^+ \mu^-}^{\tt min}$ selection. 
  Results are shown for the 3 signal scenarios of each operator: $S_{RR} = 1, T_{RR} = 1, V_{RR} = \pm 1$. See also text. 
  }
  \label{tab:selections:ttll}
\resizebox{\textwidth}{!}{
\begin{tabular}{c|c|c|c|c|c|c|c|}
\multirow{1}{*}{} \
  & \multicolumn{3}{c|}{Jet Selections: $N_j \geq 3, N_b \geq 1$} &\multicolumn{4}{c|}{Jet Selections: $N_j \geq 2, N_b \geq 1$}
  \tabularnewline
\cline{2-8} 
  & Final State & \multicolumn{2}{c|}{\boldmath{$pp \to \mu^+ \mu^- +X$}} & \multicolumn{2}{c|}{\boldmath{$pp \to \mu^+ \mu^- \ell^{\pm} +X$}} & \multicolumn{2}{c|}{\boldmath{$pp \to \mu^+ \mu^- \ell^{\pm} \ell^{'\mp} +X$}}
  \tabularnewline
\cline{2-8} 
  & Coupling & $m_{\mu^+ \mu^-}^{\tt min}$~[GeV] & $\Lambda_{\tt min}(95\%~CL)$ [TeV] & $m_{\mu^+ \mu^-}^{\tt min}$~[GeV] & $\Lambda_{\tt min}(95\%~CL)$ [TeV] & $m_{\mu^+ \mu^-}^{\tt min}$~[GeV] & $\Lambda_{\tt min}(95\%~CL)$ [TeV] 
  \tabularnewline
\hline 
\hline
\
 & $S_{RR}=1$   & & 0.8 & & $0.7$ & & $0.4$ \phantom{\noeft} \tabularnewline
\
${\cal L}=140$~fb$^{-1}$ & $T_{RR}=1$ 
 & 1400 & 1.6 & $500$ & $1.8$ & $300$ & $1.0$ \phantom{\noeft} \tabularnewline
\
& $V_{RR}=1 (-1)$
  & & $1.0~(1.0)$ & & $1.1~(1.1)$ & & $0.7~(0.7)$ \phantom{\noeft}  
  \tabularnewline
\hline 
\hline 
\
 & $S_{RR}=1$   & & 0.9 & & $0.8$ & & $0.5$ \phantom{\noeft} \tabularnewline
\
${\cal L}=300$~fb$^{-1}$ & $T_{RR}=1$ 
 & 1400 & $1.8$ & $500$ & $2.0$ & $300$ & $1.1$ \phantom{\noeft} \tabularnewline
\
& $V_{RR}=1 (-1)$
  & & $1.1 (1.1)$ & & $1.3 (1.2)$ & & $0.8 (0.7)$ \phantom{\noeft}  
  \tabularnewline
\hline 
\hline 
& $S_{RR}=1$
   &  & $1.4$ & & $1.3$ & & $0.8$ \phantom{\noeft} \tabularnewline
\
${\cal L}=3000$~fb$^{-1}$ & $T_{RR}=1$
  & 1600 & $2.8$ & $800$ & $2.9$ & $500$ & $1.8$ \phantom{\noeft} \tabularnewline
\
& $V_{RR}=1 (-1)$
  & & $1.8 (1.7)$ &  & $2.0 (1.9)$ & & $1.3 (1.2)$ \phantom{\noeft} \tabularnewline
\hline
\hline
\end{tabular}
}
\end{table*}

\subsection{Ratio observables and LFUV}

As mentioned earlier, the 4-Fermi operators of Eq.~\ref{4fermimatrix} may generate LFUV effects, in particular, asymmetric rates of the multi-lepton signals in~Eq.~\ref{lepton_chan} involving muons versus electrons, which are otherwise (i.e., within the SM) expected to be equal.

It is, therefore, useful to define generic LFU tests for multi-lepton production at the LHC, which are sensitive to the new $tt\ell\ell$ 4-Fermi interactions and which can measure the differences between muons/electrons-asymmetric final states. 
For this purpose we define the following ratio observables of cross-sections:
\begin{eqnarray}
R_{\mu/ e}^{2l} &=&
\frac{\sigma(pp \to \mu^+  \mu^- +X)}{ \sigma(pp \to e^+  e^-  + X)} ~, \label{eq:R2l} \\
R_{\mu/ e}^{3l} &=&
\frac{\sum_{\ell = e,\mu} \sigma(pp \to \ell^\pm \mu^+  \mu^-  +X)}{\sum_{\ell = e,\mu} \sigma(pp \to \ell^\pm e^+  e^-  + X)} ~, \label{eq:R3l} \\
R_{\mu /e}^{4l} &=&
\frac{\sum_{\ell,\ell^\prime = e,\mu} \sigma(pp \to \ell^{\pm} \ell^{\prime \mp} \mu^+  \mu^-  + X)}{\sum_{\ell = e,\mu} \sigma(pp \to \ell^{\pm} \ell^{\prime \mp} e^+  e^- + X)} ~, \label{eq:R4l}
\end{eqnarray}
where $\ell, \ell^\prime=e,\mu$ and
$X$ contains the accompanied jets and missing energy,
which depends on the various underlying processes that contribute to these ratios. As described  above, $X$ is different for the NP signals and for the background and this will be used here also by applying the channel-dependent jet selections of Table \ref{tab:selections} on $X$ and using, as well, a lower cut on the invariant mass of the "none-top" OSSF di-leptons to further isolate the signals from the background, as described below.

These ratio observables are particularly useful and reliable probes of LFUV NP, since they potentially minimize the effects of the theoretical uncertainties involved in the calculation of the corresponding cross-sections
(see e.g., \cite{LFU-our}) as well as the experimental systematic uncertainties.\footnote{Lepton flavor independent uncertainties from NLO QCD effects, loop corrections from EFT operators 
(see e.g., \cite{Dawson:2018dxp}) as well as PDF uncertainties are expected to be canceled to a large extent in such ratio observables. 
Even so, the impact of the theoretical uncertainties is accounted for in our analysis, as a part of the total overall uncertainties that we consider below.} 
In particular, the new $tt\mu\mu$ 4-Fermi terms contribute only to the numerators of the ratios in~Eq.~\ref{eq:R2l},~Eq.~\ref{eq:R3l} and~Eq.~\ref{eq:R4l}, thus leading to $R^{nl}_{\mu/e} \neq 1$. 
On the other hand in the SM, deviations from unity for all these ratio observables, e.g., through the non-universal Higgs-lepton Yukawa couplings to leptons, are much smaller than the expected
experimental accuracy --  as is the case, {\it in particular}, for high $p_T$ events which are our primary interest in this work. Effects of non-universal
reconstruction efficiencies and acceptance for the different leptonic final states will be included in the overall uncertainty assumed below for the measurement of $R^{nl}_{\mu/e}$  in Eq.~\ref{eq:R2l} - Eq.~\ref{eq:R4l}.

For each ratio observable in Eq.~\ref{eq:R2l} - Eq.~\ref{eq:R4l} we have (this holds also in the case of NP scenarios that are universal in lepton flavors)
\begin{eqnarray}
R^{nl}_{\mu /e} \sim 1 + \delta (\Lambda)~, \label{deltaR1}
\end{eqnarray}
where the NP effect is contained in $\delta (\Lambda)$ and, using the cumulative   
cross-section of Eq.~\ref{NP-CSX}, we have:
\begin{eqnarray}
\delta  (\Lambda) \propto \frac{c^{\tt INT} \cdot \sigma_{n \ell}^{\tt INT} + c^{\tt NP} \cdot \sigma_{n \ell}^{\tt NP}}{\sigma_{n \ell}^{\tt SM}}~, \label{deltaR2}
\end{eqnarray}
where $c^{\tt INT}$ and $c^{\tt NP}$ are the 
NP couplings of the interference and NP$^2$ terms, respectively, as given in Eq.~\ref{NP-couplings}.

To study the sensitivity to the potential LFUV signal we then define the following $\chi^2$-like test (dropping the subscript $\mu/e$ hear after):
\begin{eqnarray}
\chi^2 = \sum_{n=2,3,4} \frac{\left[ R^{nl}(\Lambda) - R^{nl}({\tt exp}) \right]^2}{\left(\delta R^{nl} \right)^2} ~,
\label{eq:chi2} 
\end{eqnarray}
where $R^{nl}({\tt exp})$ is the expected experimental measured value of the ratio (see discussion below) and  $\delta R^{nl}$ denotes the corresponding overall (experimental plus theoretical) statistic + systematic $1 \sigma$ uncertainty.

Then, based on the expectation from the corresponding  irreducible SM + background cross-sections (i.e., assuming no NP in the data), we use in our $\chi^2$-like test a lower cut ($m_{\ell \ell}^{min}$) on the invariant mass of the "none-top" OSSF di-leptons, i.e., 
$m_{\ell^+ \ell^-}^{OSSF(nt)} > m_{\ell \ell}^{\tt min}$, 
that ensures at least five event in each channel for a given luminosity,  
e.g., $(\sigma^{SM} + \sigma^{WZ}) \cdot {\cal L} > 5$ in the tri-lepton case.
The resulting channel and luminosity dependent $m_{\ell \ell}^{\tt min}$ cuts are listed in Table \ref{tab:mll-selections}. 

\begin{table}[htb]
\caption{The luminosity and channel dependent $m_{\ell \ell}^{\tt min}$
selections on the invariant mass of the "none-top" OSSF di-lepton, i.e., $m_{\ell \ell}^{OSSF(nt)} > m_{\ell \ell}^{\tt min}$, that were used in the $\chi^2$-like test of Eq.~\ref{eq:chi2}, for the different multi-lepton channels. See also text. 
\label{tab:mll-selections}}
\begin{tabular}{c|ccc|}
& \multicolumn{3}{c|}{$m_{\ell \ell}^{\tt min}$ [GeV]} 
\tabularnewline
\cline{2-4}
 ${\cal L}$  [fb$^{-1}$] & $2 \ell$ & $3 \ell$ & $4 \ell$  
\tabularnewline
\hline 
\
$140$ & 1000  & 200  & 100   \tabularnewline
$300$ &  1100  & 200 & 100  \tabularnewline
$3000$ &  1400  & 500  & 300  \tabularnewline
\hline 
\end{tabular}
\end{table}

For the purpose of extracting a bound on $ \Lambda $ we assume that, on average, no NP is observed. 
In this case we expect the experimental values 
$R^{nl}({\tt exp})$ to be normally distributed with unit mean and standard deviation $\delta R^{nl}$. Thus, if $ {\cal P}^{3l}_{\tt exp} $ denotes the  probability distribution function (PDF) for  $R^{nl}({\tt exp})$, then
\begin{eqnarray}
{\cal P}^{nl}_{\tt exp} = {\cal N} \left(1, \left(\delta R^{n \ell} \right)^2 \right) ~,  \label{eq:normal}  
\end{eqnarray}
where $ {\cal N}(a,s^2)$ denotes the normal distribution for average $a$ and standard deviation $s$. 

We don't know the actual uncertainties of the experiment and we, therefore, choose two potentially realistic benchmark values
for the overall uncertainties $\delta R^{n \ell}$ of the data samples (see e.g., \cite{CMS-dilepton}): $\delta R^{n \ell} = 15\%,25\%$ for all channels, i.e., assuming for simplicity a common overall uncertainty in the di-, tri- and four-leptons ratios. We assume that these benchmark uncertainties account for both the experimental and the theoretical systematic uncertainties, where the latter is expected to be minimized due to the use of ratio observables (see also discussion above).
Moreover, we assume in Eq.~\ref{eq:chi2} that 
the systematic uncertainties in each channel are uncorrelated, since the information about the correlation matrix of the uncertainties is not yet available for the measurements/channels used in our $\chi^2$-like test.\footnote{\label{foot.2}In the general case, where the correlation matrix for the systematic uncertainties is provided, the $\chi^2$-test 
reads instead: 
$\chi^2 = \sum_{ij} \left( R^{n\ell} (\Lambda) - R^{n \ell} (\tt exp) \right) \sigma_{n m}^{-2} \left( R^{m \ell} (\Lambda) - R^{m \ell}(\tt exp) \right)$, where $n,m$ denote the different multi-lepton channels,  
$\sigma_{nm}^{-2} = \left( \delta R^{n \ell} \rho^{n m} \delta R^{m \ell} \right)^{-1}$ and $\rho^{n m}$ is the correlation matrix provided by the experiment. 
Correlations among the systematic uncertainties in the various channels used below will degrade the sensitivity to the NP, since they effectively reduce the number of observables/channels. }

\begin{figure}[htb]
\centering
\includegraphics[width=0.52\textwidth]{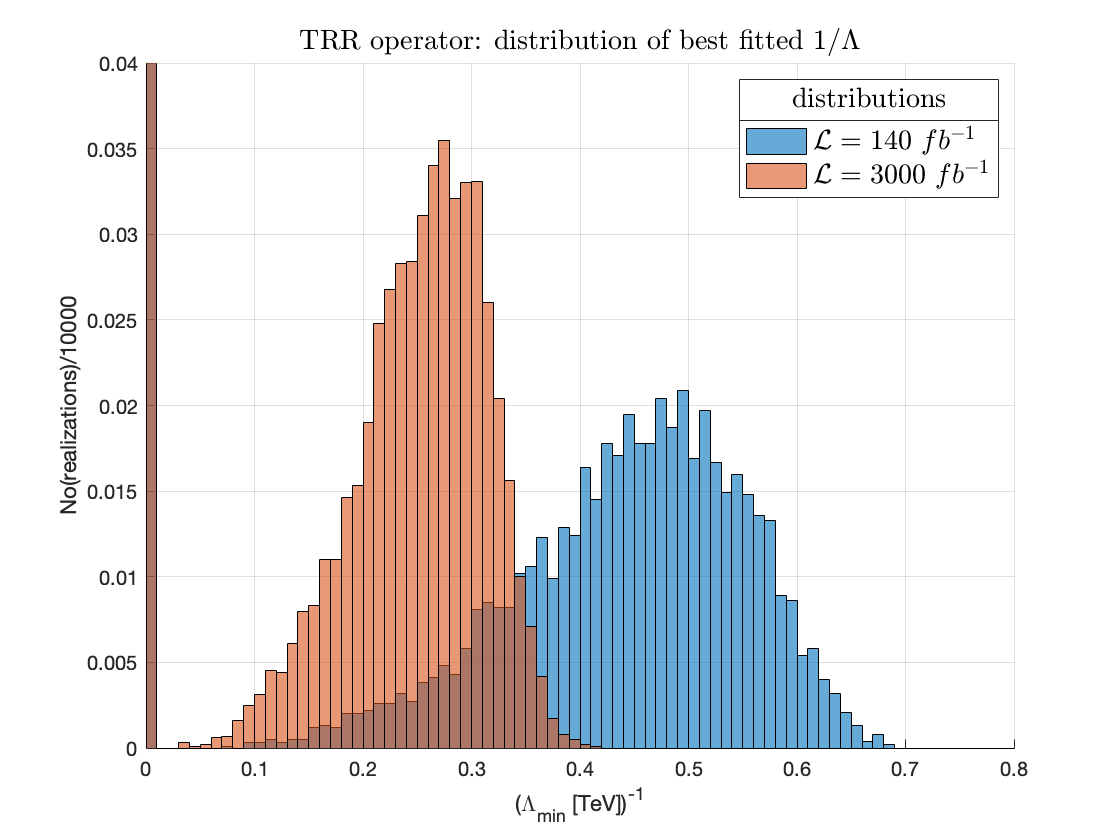}
\caption{An example of the normalized distribution of the inverse value for the best fitted $\Lambda$ of the $T_{RR}$ tensor operator, that minimizes the $\chi^2$-like test. The distributions are given for the cases ${\cal L}=140$ and $3000$~fb$^{-1}$ and for
$\delta R^{n \ell} = 50\%$. See also text.}
\centering
\label{fig:dist1}
\end{figure}

The expected bounds on $\Lambda$ are then obtained by first generating $O(10^4)$ values of $R^{nl}({\tt exp})$ distributed according to ${\cal P}^{nl}_{\tt exp}$ in Eq.~\ref{eq:normal}.
Then, for each of these $O(10^4)$ realizations of $R^{nl}({\tt exp})$ we determine the value of $ \Lambda$ that minimizes $\chi^2$ in Eq.~\ref{eq:chi2}, which we denote as $ \Lambda_{\tt min} $; the distribution of the  $ \Lambda_{\tt min} $ is also expected to be Gaussian, an example is shown in Fig.~\ref{fig:dist1}.
Finally, the expected bounds on $\Lambda$ are extracted
from this Gaussian distribution of 
$ \Lambda_{\tt min} $. 
The resulting bounds on the scale of the scalar, vector and tensor 4-Fermi operators are given in Table \ref{tab:bounds_lam} for three LHC integrated luminosity scenarios: 
${\cal L}=140,300,3000$~fb$^{-1}$, corresponding to the currently accumulated LHC luminosity, the RUN-3 projections and the planned HL-LHC luminosity, respectively. We see that the sensitivity obtained on the scale of the $tt \mu \mu$ 4-Fermi operators using our LFUV $\chi^2$-like test NP is slightly better than those obtained using the "cut and count" method of the previous chapter; this is because the $\chi^2$-like test of Eq.~\ref{eq:chi2} is using all the three multi-lepton channels, thus exploiting the fact that these channels are theoretically correlated, i.e., that the NP signals in the di-, tri- and four-leptons channels that we considered are sourced from the same underlying heavy physics - same 4-Fermi operator.   

\begin{table}[htb]
\caption{Expected bounds on the scale of the 4-Fermi
$t t \mu \mu$ with $S_{RR}=1,T_{RR}=1,V_{RR}= \pm 1$, obtained from the $\chi^2$-like test of Eq.~\ref{eq:chi2},  using the ratios 
$R^{n \ell}_{\mu/e}$ ($n=2,3,4$) of Eqs.~\ref{eq:R2l} - \ref{eq:R4l} for the di-, tri- and four-leptons signals. Results are shown for integrated luminosities of ${\cal L}=140,~300$ and $3000$~fb$^{-1}$ and for two choices of the overall uncertainty $\delta R^{n \ell} = 15\%$ and $\delta R^{n \ell} =25\%$, assuming a common uncertainty for all channels $n=2,3,4$. See also text. 
\label{tab:bounds_lam}}
\begin{tabular}{c|c|c|c|}
\multirow{1}{*}{} \
  & \multicolumn{3}{c|}{95\% bounds on $\Lambda$ [TeV]} 
\tabularnewline
\cline{2-4}
${\cal L}$ [fb$^{-1}$]  & NP $\Downarrow $ & \boldmath{$\delta R^{n \ell} = 25\%$} & \boldmath{$\delta R^{n \ell} = 15\%$} 
\tabularnewline
\hline \hline
\
 & $S_{RR}=1$   & 0.9 &  1 \tabularnewline
\cline{2-4} 
\
$140$  & $T_{RR}=1$  & 2.1  & 2.3 \tabularnewline
\cline{2-4} 
\
& $V_{RR}=1(-1)$ & 1.3(1.2) & 1.6(1.3)\tabularnewline
\hline 
\hline 
\
 & $S_{RR}=1$   & 0.9   & 1.1\tabularnewline
\cline{2-4} 
\
$300$  & $T_{RR}=1$  & 2.2   & 2.4 \tabularnewline
\cline{2-4} 
\
& $V_{RR}=1(-1)$ & 1.4(1.2)  & 1.6(1.4)\tabularnewline
\hline 
\hline 
\
 & $S_{RR}=1$ & 1.7   & 1.9\tabularnewline
\cline{2-4}
\
$3000$  & $T_{RR}=1$ & 3.7   & 4.2\tabularnewline
\cline{2-4}
\
& $V_{RR}=1(-1)$ & 2.5(2.2)  &  2.9(2.4)\tabularnewline
\hline
\hline
\end{tabular}
\end{table}

\section{Summary}

We have studied the effects of new 4-Fermi $t t \ell \ell$ flavor diagonal interactions, which can be generated from different types of underlying heavy physics containing e.g. heavy scalars and/or vectors. We showed that these higher-dimensional $t t \ell \ell$ contact interactions can lead to new high-$p_T$ events of opposite-sign same-flavor (OSSF) di-leptons in multi-lepton production accompanied by high jet-multiplicity at the LHC, in the 
di-, tri- and four-lepton channels: $pp \to t \bar t \ell^+ \ell^- \to n \cdot \ell + m \cdot j_b + p \cdot j + \missET$, where $n=2,3,4$, $m=1,2$, $p=1-4$ and $j(j_b)=$light($b$)-jet. 

We have studied in some detail the SM background to these multi-leptons signatures and showed that a very efficient separation between the NP signals and the background can be obtained with an optimal jet-multiplicity selection and, in addition, a selection of events with high invariant mass of the OSSF di-leptons, e.g., $m_{\ell^+ \ell^-}^{\tt min}(OSSF) > 500(1500)$ TeV in the tri-lepton(di-lepton) channels at the LHC RUN3 with ${\cal L}=300$~fb$^{-1}$. 

We have shown that the current bounds on the scales of the tensor $t t \ell \ell$ operator, 
$\Lambda({\rm tensor}) \gsim 1$ TeV, and the scalar and vector ones, $\Lambda({\rm scalar/vector}) \gsim 0.5$ TeV, can be improved by a factor of 2-3, using our "cut and count" analysis. For example, 95\% CL bounds of $\Lambda \gsim 1.8(1.1) $~TeV are expected on the scale of a tensor (vector) $t t \mu \mu$ interaction, already with the current $\sim 140$~fb$^{-1}$ of LHC 
data, via the tri-lepton $pp \to \ell^{\pm} \mu^+ \mu^- + 2 \cdot j_b + 2 \cdot j + \missET$ signal; this is an improvement by a factor 
of $\sim 2$ with respect to the current bounds on these operators. 
The expected reach at the HL-LHC with 3000~fb$^{-1}$ of data is $\Lambda \gsim 3(2)$~TeV 
for the tensor (vector) $t t \ell \ell$ operators.

We furthermore explored potential searches for lepton flavor universal universality violation (LFUV) effects from the higher dimensional $tt \ell \ell$ 4-Fermi operators, that can be detected via our multi-lepton signals. In particular, we have defined ratio observables for all three di-, tri- and four-lepton channels, which 
can be used to search for new LFUV effects via a $\chi^2$-like test that exploits the theoretical correlation between the LFUV signals in these three multi-lepton channels.  
We find that the sensitivity to the scale of the LFUV $tt \ell \ell$ operators is comparable to that obtained with the "cut and count" search, using our $\chi^2$-like test. For example, 
it is possible to obtain 95\% CL bounds on the LFUV tensor $tt \ell \ell$ interactions of $\Lambda \gsim 2$ TeV  with the current LHC data of ${\cal L} = 140$~fb$^{-1}$ and of $\Lambda \gsim 3.5$ TeV
at the HL-LHC with 3000~fb$^{-1}$. 

Finally, we believe that the sensitivity obtained in this paper to the new $tt \ell \ell$ 4-Fermi operators using the di-, tri- and four-leptons signals from the underlying $pp \to t \bar t \ell^+ \ell^-$ process, can be improved with further optimization of the selections that isolate the NP signals in these channels from the SM background.

\acknowledgments
The work of AS  was supported in part by the U.S. DOE contract \#DE-SC0012704.

\bibliographystyle{hunsrt.bst}
\bibliography{mybib2}

\begin{thebibliography}{100}

\bibitem{FCtopdecay1}
G.~Eilam, J.L. Hewett, and A.~Soni.
\newblock {Rare decays of the top quark in the standard and two Higgs doublet
  models}.
\newblock {\em Phys. Rev. D}, 44:1473--1484, 1991.
\newblock [Erratum: Phys.Rev.D 59, 039901 (1999)].

\bibitem{FCtopdecay2}
W.~Buchmuller and M.~Gronau.
\newblock {FLAVOR CHANGING Z0 DECAYS}.
\newblock {\em Phys. Lett. B}, 220:641--645, 1989.

\bibitem{FCtopdecay3}
Harald Fritzsch.
\newblock {T Quarks May Decay Into $Z$ Bosons and Charm}.
\newblock {\em Phys. Lett. B}, 224:423--425, 1989.

\bibitem{FCtopdecay4}
J.L. Diaz-Cruz, R.~Martinez, M.A. Perez, and A.~Rosado.
\newblock {Flavor Changing Radiative Decay of Thf T Quark}.
\newblock {\em Phys. Rev. D}, 41:891--894, 1990.

\bibitem{FCtopdecay5}
B.~Dutta-Roy, B.A. Irwin, B.~Margolis, J.~Robinson, H.D. Trottier, and
  C.~Hamazaoui.
\newblock {Threshold enhancement and the flavor changing electromagnetic
  vertex}.
\newblock {\em Phys. Rev. Lett.}, 65:827--830, 1990.

\bibitem{FCtopdecay6}
J.L. Diaz-Cruz and G.~Lopez~Castro.
\newblock {CP violation and FCNC with the top quark}.
\newblock {\em Phys. Lett. B}, 301:405--408, 1993.

\bibitem{FCtopdecay7}
B.~Mele, S.~Petrarca, and A.~Soddu.
\newblock {A New evaluation of the t ---\ensuremath{>} cH decay width in the
  standard model}.
\newblock {\em Phys. Lett. B}, 435:401--406, 1998, hep-ph/9805498.

\bibitem{FCtopdecay8}
Itzhak Baum, Gad Eilam, and Shaouly Bar-Shalom.
\newblock {Scalar flavor changing neutral currents and rare top quark decays in
  a two H iggs doublet model 'for the top quark'}.
\newblock {\em Phys. Rev. D}, 77:113008, 2008, 0802.2622.

\bibitem{t_to_cdecay_soni}
Kaustubh Agashe, Gilad Perez, and Amarjit Soni.
\newblock {Flavor structure of warped extra dimension models}.
\newblock {\em Phys. Rev. D}, 71:016002, 2005, hep-ph/0408134.

\bibitem{t_to_c_Hou}
Kai-Feng Chen, Wei-Shu Hou, Chung Kao, and Masaya Kohda.
\newblock {When the Higgs meets the Top: Search for $t \to ch^0$ at the LHC}.
\newblock {\em Phys. Lett. B}, 725:378--381, 2013, 1304.8037.

\bibitem{FCtopprod1}
Alan Axelrod.
\newblock {Flavor Changing Z0 Decay and the Top Quark}.
\newblock {\em Nucl. Phys. B}, 209:349--371, 1982.

\bibitem{FCtopprod2}
M.~Clements, C.~Footman, Andreas~S. Kronfeld, S.~Narasimhan, and D.~Photiadis.
\newblock {Flavor Changing Decays of the Z0}.
\newblock {\em Phys. Rev. D}, 27:570, 1983.

\bibitem{FCtopprod3}
Gad Eilam.
\newblock {Production of a Single Heavy Quark in $e^+ e^-$ Collisions}.
\newblock {\em Phys. Rev. D}, 28:1202, 1983.

\bibitem{FCtopprod4}
Chao-Hsi Chang, Xue-Qian Li, Jian-Xiong Wang, and Mao-Zhi Yang.
\newblock {The Production of t anti-c or anti-t c quark pair by e+ e- collision
  based on the standard model and its extensions}.
\newblock {\em Phys. Lett. B}, 313:389--394, 1993.

\bibitem{FCtopprod5}
Chao-Shang Huang, Xiao-Hong Wu, and Shou-Hua Zhu.
\newblock {Top charm associated production at high-energy e+ e- colliders in
  standard model}.
\newblock {\em Phys. Lett. B}, 452:143--149, 1999, hep-ph/9901369.

\bibitem{eetc_soni1}
David Atwood, Laura Reina, and Amarjit Soni.
\newblock {Probing flavor changing top - charm - scalar interactions in $e^{+}
  e^{-}$ collisions}.
\newblock {\em Phys. Rev. D}, 53:1199--1201, 1996, hep-ph/9506243.

\bibitem{eetc_soni2}
Kaustubh Agashe, Gilad Perez, and Amarjit Soni.
\newblock {Collider Signals of Top Quark Flavor Violation from a Warped Extra
  Dimension}.
\newblock {\em Phys. Rev. D}, 75:015002, 2007, hep-ph/0606293.

\bibitem{eetc_Hou1}
Wei-Shu Hou, Guey-Lin Lin, and Chien-Yi Ma.
\newblock {Flavor changing neutral Higgs couplings and top charm production at
  next linear collider}.
\newblock {\em Phys. Rev. D}, 56:7434--7443, 1997, hep-ph/9708228.

\bibitem{ourreview}
David Atwood, Shaouly Bar-Shalom, Gad Eilam, and Amarjit Soni.
\newblock {CP violation in top physics}.
\newblock {\em Phys. Rept.}, 347:1--222, 2001, hep-ph/0006032.

\bibitem{Degrande:2018fog}
Celine Degrande, Fabio Maltoni, Ken Mimasu, Eleni Vryonidou, and Cen Zhang.
\newblock {Single-top associated production with a $Z$ or $H$ boson at the LHC:
  the SMEFT interpretation}.
\newblock {\em JHEP}, 10:005, 2018, 1804.07773.

\bibitem{Maltoni-global}
Gauthier Durieux, Fabio Maltoni, and Cen Zhang.
\newblock {Global approach to top-quark flavor-changing interactions}.
\newblock {\em Phys. Rev. D}, 91(7):074017, 2015, 1412.7166.

\bibitem{Hartland:2019bjb}
Nathan~P. Hartland, Fabio Maltoni, Emanuele~R. Nocera, Juan Rojo, Emma Slade,
  Eleni Vryonidou, and Cen Zhang.
\newblock {A Monte Carlo global analysis of the Standard Model Effective Field
  Theory: the top quark sector}.
\newblock {\em JHEP}, 04:100, 2019, 1901.05965.

\bibitem{Durieux:2019rbz}
Gauthier Durieux, Adrian Irles, Víctor Miralles, Ana Peñuelas, Roman Pöschl,
  Martín Perelló, and Marcel Vos.
\newblock {The electro-weak couplings of the top and bottom quarks -- global
  fit and future prospects}.
\newblock {\em JHEP}, 12:098, 2019, 1907.10619.

\bibitem{singletop1}
Timothy~M.P. Tait and C.-P. Yuan.
\newblock {Single top quark production as a window to physics beyond the
  standard model}.
\newblock {\em Phys. Rev. D}, 63:014018, 2000, hep-ph/0007298.

\bibitem{tZgamma1}
Nikolaos Kidonakis and Alexander Belyaev.
\newblock {FCNC top quark production via anomalous tqV couplings beyond leading
  order}.
\newblock {\em JHEP}, 12:004, 2003, hep-ph/0310299.

\bibitem{SMEFTtop1}
Ilaria Brivio, Sebastian Bruggisser, Fabio Maltoni, Rhea Moutafis, Tilman
  Plehn, Eleni Vryonidou, Susanne Westhoff, and C.~Zhang.
\newblock {O new physics, where art thou? A global search in the top sector}.
\newblock {\em JHEP}, 02:131, 2020, 1910.03606.

\bibitem{tZ3}
D.~Barducci et~al.
\newblock {Interpreting top-quark LHC measurements in the standard-model
  effective field theory}.
\newblock 2 2018, 1802.07237.

\bibitem{SMEFTtop2}
Fabio Maltoni, Luca Mantani, and Ken Mimasu.
\newblock {Top-quark electroweak interactions at high energy}.
\newblock {\em JHEP}, 10:004, 2019, 1904.05637.

\bibitem{SMEFTtop3}
Tobias Neumann and Zack~Edward Sullivan.
\newblock {Off-Shell Single-Top-Quark Production in the Standard Model
  Effective Field Theory}.
\newblock {\em JHEP}, 06:022, 2019, 1903.11023.

\bibitem{SMEFTtop4}
Christoph Englert, Michael Russell, and Chris~D. White.
\newblock {Effective Field Theory in the top sector: do multijets help?}
\newblock {\em Phys.\ Rev.\ D}, 99(3):035019, 2019, 1809.09744.

\bibitem{Zhang:2016omx}
Cen Zhang.
\newblock {Single Top Production at Next-to-Leading Order in the Standard Model
  Effective Field Theory}.
\newblock {\em Phys. Rev. Lett.}, 116(16):162002, 2016, 1601.06163.

\bibitem{tH1}
F.~Maltoni, K.~Paul, T.~Stelzer, and S.~Willenbrock.
\newblock {Associated production of Higgs and single top at hadron colliders}.
\newblock {\em Phys. Rev. D}, 64:094023, 2001, hep-ph/0106293.

\bibitem{tH2}
Sanjoy Biswas, Emidio Gabrielli, and Barbara Mele.
\newblock {Single top and Higgs associated production as a probe of the Htt
  coupling sign at the LHC}.
\newblock {\em JHEP}, 01:088, 2013, 1211.0499.

\bibitem{tH3}
Marco Farina, Christophe Grojean, Fabio Maltoni, Ennio Salvioni, and Andrea
  Thamm.
\newblock {Lifting degeneracies in Higgs couplings using single top production
  in association with a Higgs boson}.
\newblock {\em JHEP}, 05:022, 2013, 1211.3736.

\bibitem{tH4}
Federico Demartin, Fabio Maltoni, Kentarou Mawatari, and Marco Zaro.
\newblock {Higgs production in association with a single top quark at the LHC}.
\newblock {\em Eur. Phys. J. C}, 75(6):267, 2015, 1504.00611.

\bibitem{tH5}
Lei Wu.
\newblock {Enhancing $thj$ Production from Top-Higgs FCNC Couplings}.
\newblock {\em JHEP}, 02:061, 2015, 1407.6113.

\bibitem{tH6}
Wei Liu, Hao Sun, XiaoJuan Wang, and Xuan Luo.
\newblock {Probing the anomalous FCNC top-Higgs Yukawa couplings at the Large
  Hadron Electron Collider}.
\newblock {\em Phys. Rev. D}, 92(7):074015, 2015, 1507.03264.

\bibitem{tH7}
Yao-Bei Liu and Zhen-Jun Xiao.
\newblock {Searches for the FCNC couplings from top-Higgs associated production
  signal with $h\to \gamma\gamma$ at the LHC}.
\newblock {\em Phys. Lett. B}, 763:458--464, 2016, 1610.03250.

\bibitem{tH8}
Yao-Bei Liu and Zhen-Jun Xiao.
\newblock {Searches for top-Higgs FCNC couplings via the Whj signal with $h\to
  \gamma\gamma$ at the LHC}.
\newblock {\em Phys. Rev. D}, 94(5):054018, 2016, 1605.01179.

\bibitem{tZ1}
Jie-Fen Shen, Yu-Qi Li, and Yao-Bei Liu.
\newblock {Searches for anomalous tqZ couplings from the trilepton signal of tZ
  associated production at the 14 TeV LHC}.
\newblock {\em Phys. Lett. B}, 776:391--395, 2018, 1712.03506.

\bibitem{tZ2}
F.~del Aguila, J.A. Aguilar-Saavedra, and L.~Ametller.
\newblock {Z t and gamma t production via top flavor changing neutral couplings
  at the Fermilab Tevatron}.
\newblock {\em Phys. Lett. B}, 462:310--318, 1999, hep-ph/9906462.

\bibitem{tZ4}
Hamzeh Khanpour, Sara Khatibi, Morteza Khatiri~Yanehsari, and Mojtaba
  Mohammadi~Najafabadi.
\newblock {Single top quark production as a probe of anomalous $tq\gamma$ and
  $tqZ$ couplings at the FCC-ee}.
\newblock {\em Phys. Lett. B}, 775:25--31, 2017, 1408.2090.

\bibitem{tZ5}
Nikolaos Kidonakis.
\newblock {Higher-order corrections for $tZ$ production via anomalous
  couplings}.
\newblock {\em Phys. Rev. D}, 97(3):034028, 2018, 1712.01144.

\bibitem{tZ6}
P.M. Ferreira, R.B. Guedes, and R.~Santos.
\newblock {Combined effects of strong and electroweak FCNC effective operators
  in top quark physics at the CERN LHC}.
\newblock {\em Phys. Rev. D}, 77:114008, 2008, 0802.2075.

\bibitem{tZ7}
Bo~Hua Li, Yue Zhang, Chong~Sheng Li, Jun Gao, and Hua~Xing Zhu.
\newblock {Next-to-leading order QCD corrections to $tZ$ associated production
  via the flavor-changing neutral-current couplings at hadron colliders}.
\newblock {\em Phys. Rev. D}, 83:114049, 2011, 1103.5122.

\bibitem{tZ8-trilepton}
Jean-Laurent Agram, Jeremy Andrea, Eric Conte, Benjamin Fuks, Denis Gel\'e, and
  Pierre Lansonneur.
\newblock {Probing top anomalous couplings at the LHC with trilepton signatures
  in the single top mode}.
\newblock {\em Phys. Lett. B}, 725:123--126, 2013, 1304.5551.

\bibitem{tZ9-trilepton}
Yao-Bei Liu and Stefano Moretti.
\newblock {Probing tqZ anomalous couplings in the trilepton signal at the
  HL-LHC, HE-LHC and FCC-hh}.
\newblock 10 2020, 2010.05148.

\bibitem{tgamma1}
Yu-Chen Guo, Chong-Xing Yue, and Shuo Yang.
\newblock {Search for anomalous couplings via single top quark production in
  association with a photon at LHC}.
\newblock {\em Eur. Phys. J. C}, 76(11):596, 2016, 1603.00604.

\bibitem{tZEFT-decay1}
Tao Han, R.D. Peccei, and X.~Zhang.
\newblock {Top quark decay via flavor changing neutral currents at hadron
  colliders}.
\newblock {\em Nucl. Phys. B}, 454:527--540, 1995, hep-ph/9506461.

\bibitem{tZprime}
Ezequiel Alvarez, Aurelio Juste, Manuel Szewc, and Tamara Vazquez~Schroeder.
\newblock {Topping-up multilepton plus b-jets anomalies at the LHC with a $Z'$
  boson}.
\newblock 11 2020, 2011.06514.

\bibitem{Gudron1}
Rigo Bause, Hector Gisbert, Marcel Golz, and Gudrun Hiller.
\newblock {Lepton universality and lepton flavor conservation tests with
  dineutrino modes}.
\newblock {\em Eur. Phys. J. C}, 82(2):164, 2022, 2007.05001.

\bibitem{Gudron2}
Stefan Bi\ss{}mann, Cornelius Grunwald, Gudrun Hiller, and Kevin Kr\"oninger.
\newblock {Top and Beauty synergies in SMEFT-fits at present and future
  colliders}.
\newblock {\em JHEP}, 06:010, 2021, 2012.10456.

\bibitem{Ellis:2020unq}
John Ellis, Maeve Madigan, Ken Mimasu, Veronica Sanz, and Tevong You.
\newblock {Top, Higgs, Diboson and Electroweak Fit to the Standard Model
  Effective Field Theory}.
\newblock {\em JHEP}, 04:279, 2021, 2012.02779.

\bibitem{Ethier:2021bye}
Jacob~J. Ethier, Fabio Maltoni, Luca Mantani, Emanuele~R. Nocera, Juan Rojo,
  Emma Slade, Eleni Vryonidou, and Cen Zhang.
\newblock {Combined SMEFT interpretation of Higgs, diboson, and top quark data
  from the LHC}.
\newblock 4 2021, 2105.00006.

\bibitem{bsll-our}
Yoav Afik, Jonathan Cohen, Eitan Gozani, Enrique Kajomovitz, and Yoram Rozen.
\newblock {Establishing a Search for $b \rightarrow s \ell^{+} \ell^{-}$
  Anomalies at the LHC}.
\newblock {\em JHEP}, 08:056, 2018, 1805.11402.

\bibitem{bbll-our}
Yoav Afik, Shaouly Bar-Shalom, Jonathan Cohen, and Yoram Rozen.
\newblock {Searching for New Physics with $b\bar{b} \ell^+ \ell^-$ contact
  interactions}.
\newblock {\em Phys. Lett. B}, 807:135541, 2020, 1912.00425.

\bibitem{LFU-our}
Yoav Afik, Shaouly Bar-Shalom, Jonathan Cohen, Amarjit Soni, and Jose Wudka.
\newblock {High $p_T$ correlated tests of lepton universality in lepton(s) +
  jet(s) processes; An EFT analysis}.
\newblock {\em Phys. Lett. B}, 811:135908, 2020, 2005.06457.

\bibitem{tull-our}
Yoav Afik, Shaouly Bar-Shalom, Amarjit Soni, and Jose Wudka.
\newblock {New flavor physics in di- and tri-lepton events from single-top at
  the LHC and beyond}.
\newblock 1 2021, 2101.05286.

\bibitem{Tonero:2020zcy}
Daniel Stolarski and Alberto Tonero.
\newblock {Constraining New Physics with Single Top production at LHC}.
\newblock 2020, 2004.07856.

\bibitem{topdecay1}
Radja Boughezal, Chien-Yi Chen, Frank Petriello, and Daniel Wiegand.
\newblock {Top quark decay at next-to-leading order in the Standard Model
  Effective Field Theory}.
\newblock {\em Phys.\ Rev.\ D}, 100(5):056023, 2019, 1907.00997.

\bibitem{1008.3562}
J.A. Aguilar-Saavedra.
\newblock {Effective four-fermion operators in top physics: A Roadmap}.
\newblock {\em Nucl. Phys. B}, 843:638--672, 2011, 1008.3562.
\newblock [Erratum: Nucl.Phys.B 851, 443--444 (2011)].

\bibitem{topdecay3}
Sacha Davidson, Michelangelo~L. Mangano, Stephane Perries, and Viola Sordini.
\newblock {Lepton Flavour Violating top decays at the LHC}.
\newblock {\em Eur. Phys. J. C}, 75(9):450, 2015, 1507.07163.

\bibitem{topdecay2}
Mikael Chala, Jose Santiago, and Michael Spannowsky.
\newblock {Constraining four-fermion operators using rare top decays}.
\newblock {\em JHEP}, 04:014, 2019, 1809.09624.

\bibitem{Gottardo:2019lmv}
Carlo~Alberto Gottardo.
\newblock {\em {Search for charged lepton-flavour violation in top-quark decays
  at the LHC with the ATLAS detector}}.
\newblock PhD thesis, U. Bonn (main), 2019.

\bibitem{LHCb:2021trn}
Roel Aaij et~al.
\newblock {Test of lepton universality in beauty-quark decays}.
\newblock 3 2021, 2103.11769.

\bibitem{deSimone:2020kwi}
P.~de~Simone.
\newblock {Experimental Review on Lepton Universality and Lepton Flavour
  Violation tests in $B$ decays}.
\newblock {\em EPJ Web Conf.}, 234:01004, 2020.

\bibitem{Gherardi:2019zil}
Valerio Gherardi, David Marzocca, Marco Nardecchia, and Andrea Romanino.
\newblock {Rank-One Flavor Violation and B-meson anomalies}.
\newblock {\em JHEP}, 10:112, 2019, 1903.10954.

\bibitem{Aaij:2014pli}
R.~Aaij et~al.
\newblock {Differential branching fractions and isospin asymmetries of $B \to
  K^{(*)} \mu^+ \mu^-$ decays}.
\newblock {\em JHEP}, 06:133, 2014, 1403.8044.

\bibitem{Aaij:2014ora}
Roel Aaij et~al.
\newblock {Test of lepton universality using $B^{+}\rightarrow
  K^{+}\ell^{+}\ell^{-}$ decays}.
\newblock {\em Phys. Rev. Lett.}, 113:151601, 2014, 1406.6482.

\bibitem{Aaij:2017vbb}
R.~Aaij et~al.
\newblock {Test of lepton universality with $B^{0} \rightarrow
  K^{*0}\ell^{+}\ell^{-}$ decays}.
\newblock {\em JHEP}, 08:055, 2017, 1705.05802.

\bibitem{Aaij:2015esa}
Roel Aaij et~al.
\newblock {Angular analysis and differential branching fraction of the decay
  $B^0_s\to\phi\mu^+\mu^-$}.
\newblock {\em JHEP}, 09:179, 2015, 1506.08777.

\bibitem{Aaij:2015oid}
Roel Aaij et~al.
\newblock {Angular analysis of the $B^{0} \to K^{*0} \mu^{+} \mu^{-}$ decay
  using 3 fb$^{-1}$ of integrated luminosity}.
\newblock {\em JHEP}, 02:104, 2016, 1512.04442.

\bibitem{Wehle:2016yoi}
S.~Wehle et~al.
\newblock {Lepton-Flavor-Dependent Angular Analysis of $B\to K^\ast
  \ell^+\ell^-$}.
\newblock {\em Phys. Rev. Lett.}, 118(11):111801, 2017, 1612.05014.

\bibitem{Abdesselam:2016llu}
A.~Abdesselam et~al.
\newblock {Angular analysis of $B^0 \to K^\ast(892)^0 \ell^+ \ell^-$}.
\newblock In {\em {Proceedings, LHCSki 2016 - A First Discussion of 13 TeV
  Results: Obergurgl, Austria, April 10-15, 2016}}, 2016, 1604.04042.

\bibitem{ATLAS:2017dlm}
The~ATLAS collaboration.
\newblock {Angular analysis of $B^0_d \to K^{*}\mu^+\mu^-$ decays in $pp$
  collisions at $\sqrt{s}= 8$ TeV with the ATLAS detector}.
\newblock 2017.

\bibitem{CMS:2017ivg}
CMS Collaboration.
\newblock {Measurement of the $P_1$ and $P_5'$ angular parameters of the decay
  $\mathrm{B}^0 \to \mathrm{K}^{*0} \mu^+ \mu^-$ in proton-proton collisions at
  $\sqrt{s}=8~\mathrm{TeV}$}.
\newblock 2017.

\bibitem{Bifani:2017gyn}
Simone Bifani.
\newblock {Status of New Physics searches with $b \to s \ell^{+}\ell^{-}$
  transitions @ LHCb}.
\newblock In {\em {Proceedings, 52nd Rencontres de Moriond on Electroweak
  Interactions and Unified Theories: La Thuile, Italy, March 18-25, 2017}},
  pages 197--202, 2017, 1705.02693.

\bibitem{Aaij:2019wad}
Roel Aaij et~al.
\newblock {Search for lepton-universality violation in $B^+\to K^+\ell^+\ell^-$
  decays}.
\newblock {\em Phys. Rev. Lett.}, 122(19):191801, 2019, 1903.09252.

\bibitem{Abdesselam:2019wac}
A.~Abdesselam et~al.
\newblock {Test of lepton flavor universality in ${B\to K^\ast\ell^+\ell^-}$
  decays at Belle}.
\newblock 2019, 1904.02440.

\bibitem{Lees:2012xj}
J.~P. Lees et~al.
\newblock {Evidence for an excess of $\bar{B} \to D^{(*)} \tau^-\bar{\nu}_\tau$
  decays}.
\newblock {\em Phys. Rev. Lett.}, 109:101802, 2012, 1205.5442.

\bibitem{Lees:2013uzd}
J.~P. Lees et~al.
\newblock {Measurement of an Excess of $\bar{B} \to D^{(*)}\tau^-
  \bar{\nu}_\tau$ Decays and Implications for Charged Higgs Bosons}.
\newblock {\em Phys. Rev.}, D88(7):072012, 2013, 1303.0571.

\bibitem{Huschle:2015rga}
M.~Huschle et~al.
\newblock {Measurement of the branching ratio of $\bar{B} \to D^{(\ast)} \tau^-
  \bar{\nu}_\tau$ relative to $\bar{B} \to D^{(\ast)} \ell^- \bar{\nu}_\ell$
  decays with hadronic tagging at Belle}.
\newblock {\em Phys. Rev.}, D92(7):072014, 2015, 1507.03233.

\bibitem{Hirose:2016wfn}
S.~Hirose et~al.
\newblock {Measurement of the $\tau$ lepton polarization and $R(D^*)$ in the
  decay $\bar{B} \to D^* \tau^- \bar{\nu}_\tau$}.
\newblock {\em Phys. Rev. Lett.}, 118(21):211801, 2017, 1612.00529.

\bibitem{Aaij:2015yra}
Roel Aaij et~al.
\newblock {Measurement of the ratio of branching fractions
  $\mathcal{B}(\bar{B}^0 \to
  D^{*+}\tau^{-}\bar{\nu}_{\tau})/\mathcal{B}(\bar{B}^0 \to
  D^{*+}\mu^{-}\bar{\nu}_{\mu})$}.
\newblock {\em Phys. Rev. Lett.}, 115(11):111803, 2015, 1506.08614.
\newblock [Erratum: Phys. Rev. Lett.115,no.15,159901(2015)].

\bibitem{Aaij:2017uff}
R.~Aaij et~al.
\newblock {Measurement of the ratio of the $B^0 \to D^{*-} \tau^+ \nu_{\tau}$
  and $B^0 \to D^{*-} \mu^+ \nu_{\mu}$ branching fractions using three-prong
  $\tau$-lepton decays}.
\newblock {\em Phys. Rev. Lett.}, 120(17):171802, 2018, 1708.08856.

\bibitem{Aaij:2017deq}
R.~Aaij et~al.
\newblock {Test of Lepton Flavor Universality by the measurement of the $B^0
  \to D^{*-} \tau^+ \nu_{\tau}$ branching fraction using three-prong $\tau$
  decays}.
\newblock {\em Phys. Rev.}, D97(7):072013, 2018, 1711.02505.

\bibitem{Adamczyk:2019wyt}
Karol Adamczyk.
\newblock {Semitauonic $B$ decays at Belle/Belle II}.
\newblock In {\em {10th International Workshop on the CKM Unitarity Triangle
  (CKM 2018) Heidelberg, Germany, September 17-21, 2018}}, 2019, 1901.06380.

\bibitem{Abdesselam:2019dgh}
A.~Abdesselam et~al.
\newblock {Measurement of $\mathcal{R}(D)$ and $\mathcal{R}(D^{\ast})$ with a
  semileptonic tagging method}.
\newblock 2019, 1904.08794.

\bibitem{Bifani:2018zmi}
Simone Bifani, Sébastien Descotes-Genon, Antonio Romero~Vidal, and
  Marie-Hélène Schune.
\newblock {Review of Lepton Universality tests in $B$ decays}.
\newblock {\em J.\ Phys.\ G}, 46(2):023001, 2019, 1809.06229.

\bibitem{Glashow:2014iga}
Sheldon~L. Glashow, Diego Guadagnoli, and Kenneth Lane.
\newblock {Lepton Flavor Violation in $B$ Decays?}
\newblock {\em Phys. Rev. Lett.}, 114:091801, 2015, 1411.0565.

\bibitem{gm2-recent}
B.~Abi et~al.
\newblock {Measurement of the Positive Muon Anomalous Magnetic Moment to 0.46
  ppm}.
\newblock {\em Phys. Rev. Lett.}, 126(14):141801, 2021, 2104.03281.

\bibitem{ATLAS-dilepton}
Georges Aad et~al.
\newblock {Search for new non-resonant phenomena in high-mass dilepton final
  states with the ATLAS detector}.
\newblock {\em JHEP}, 11:005, 2020, 2006.12946.
\newblock [Erratum: JHEP 04, 142 (2021)].

\bibitem{ATLAS-bsll}
Georges Aad et~al.
\newblock {Search for New Phenomena in Final States with Two Leptons and One or
  No $b$-Tagged Jets at $\sqrt{s} = 13$ TeV Using the ATLAS Detector}.
\newblock {\em Phys. Rev. Lett.}, 127(14):141801, 2021, 2105.13847.

\bibitem{CMS-dilepton}
Albert~M Sirunyan et~al.
\newblock {Search for resonant and nonresonant new phenomena in high-mass
  dilepton final states at $\sqrt{s} = $ 13 TeV}.
\newblock 3 2021, 2103.02708.

\bibitem{Fajfer:2021cxa}
Svjetlana Fajfer, Jernej~F. Kamenik, and Michele Tammaro.
\newblock {Interplay of New Physics effects in $(g-2)_{\ell}$ and $h \to \ell^+
  \ell^-$ - lessons from SMEFT}.
\newblock {\em JHEP}, 06:099, 2021, 2103.10859.

\bibitem{Aebischer:2021uvt}
Jason Aebischer, Wouter Dekens, Elizabeth~E. Jenkins, Aneesh~V. Manohar, Dipan
  Sengupta, and Peter Stoffer.
\newblock {Effective field theory interpretation of lepton magnetic and
  electric dipole moments}.
\newblock 2 2021, 2102.08954.

\bibitem{Kamenik:2017tnu}
Jernej~F. Kamenik, Yotam Soreq, and Jure Zupan.
\newblock {Lepton flavor universality violation without new sources of quark
  flavor violation}.
\newblock {\em Phys. Rev. D}, 97(3):035002, 2018, 1704.06005.

\bibitem{Fox:2018ldq}
Patrick~J. Fox, Ian Low, and Yue Zhang.
\newblock {Top-philic $Z'$ forces at the LHC}.
\newblock {\em JHEP}, 03:074, 2018, 1801.03505.

\bibitem{Camargo-Molina:2018cwu}
Jos\'e~Eliel Camargo-Molina, Alejandro Celis, and Darius~A. Faroughy.
\newblock {Anomalies in Bottom from new physics in Top}.
\newblock {\em Phys. Lett. B}, 784:284--293, 2018, 1805.04917.

\bibitem{Ciuchini:2019usw}
Marco Ciuchini, Ant\'onio~M. Coutinho, Marco Fedele, Enrico Franco, Ayan Paul,
  Luca Silvestrini, and Mauro Valli.
\newblock {New Physics in $b \to s \ell^+ \ell^-$ confronts new data on Lepton
  Universality}.
\newblock {\em Eur. Phys. J. C}, 79(8):719, 2019, 1903.09632.

\bibitem{Soni:1973pyl}
A.~Soni.
\newblock {Radiative corrections to $p + p \to \ell^+ + \ell^- +$ anything and
  application to muon-electron symmetry}.
\newblock {\em Phys. Rev. D}, 8:2264--2267, 1973.

\bibitem{our_tc_paper}
S.~Bar-Shalom and J.~Wudka.
\newblock {Flavor changing single top quark production channels at e+ e-
  colliders in the effective Lagrangian description}.
\newblock {\em Phys. Rev.}, D60:094016, 1999, hep-ph/9905407.

\bibitem{Grzadkowski:1995te}
Bohdan Grzadkowski.
\newblock {Four Fermi effective operators at $e^{+} e^{-} \to \bar{t} t$}.
\newblock {\em Acta Phys. Polon. B}, 27:921--932, 1996, hep-ph/9511279.

\bibitem{Grzadkowski:1997cj}
Bohdan Grzadkowski, Zenro Hioki, and Michal Szafranski.
\newblock {Four Fermi effective operators in top quark production and decay}.
\newblock pages 1113--1135, 12 1997, hep-ph/9712357.

\bibitem{Marzocca}
Admir Greljo and David Marzocca.
\newblock {High-$p_T$ dilepton tails and flavor physics}.
\newblock {\em Eur. Phys. J.}, C77(8):548, 2017, 1704.09015.

\bibitem{Admir}
Darius~A. Faroughy, Admir Greljo, and Jernej~F. Kamenik.
\newblock {Confronting lepton flavor universality violation in B decays with
  high-$p_T$ tau lepton searches at LHC}.
\newblock {\em Phys. Lett.}, B764:126--134, 2017, 1609.07138.

\bibitem{Marzocca:2020ueu}
David Marzocca, Ui~Min, and Minho Son.
\newblock {Bottom-Flavored Mono-Tau Tails at the LHC}.
\newblock 8 2020, 2008.07541.

\bibitem{Panico:2021vav}
Giuliano Panico, Lorenzo Ricci, and Andrea Wulzer.
\newblock {High-energy EFT probes with fully differential Drell-Yan
  measurements}.
\newblock {\em JHEP}, 07:086, 2021, 2103.10532.

\bibitem{Crivellin:2021rbf}
Andreas Crivellin, Claudio~Andrea Manzari, and Marc Montull.
\newblock {Correlating Non-Resonant Di-Electron Searches at the LHC to the
  Cabibbo-Angle Anomaly and Lepton Flavour Universality Violation}.
\newblock 3 2021, 2103.12003.

\bibitem{Greljo:2021kvv}
Admir Greljo, Shayan Iranipour, Zahari Kassabov, Maeve Madigan, James Moore,
  Juan Rojo, Maria Ubiali, and Cameron Voisey.
\newblock {Parton distributions in the SMEFT from high-energy Drell-Yan tails}.
\newblock {\em JHEP}, 07:122, 2021, 2104.02723.

\bibitem{ttV_1}
Morad Aaboud et~al.
\newblock {Measurement of the $t\bar{t}Z$ and $t\bar{t}W$ cross sections in
  proton-proton collisions at $\sqrt{s}=13$ TeV with the ATLAS detector}.
\newblock {\em Phys. Rev. D}, 99(7):072009, 2019, 1901.03584.

\bibitem{ttV_2}
Albert~M Sirunyan et~al.
\newblock {Measurement of the cross section for top quark pair production in
  association with a W or Z boson in proton-proton collisions at $\sqrt{s} =$
  13 TeV}.
\newblock {\em JHEP}, 08:011, 2018, 1711.02547.

\bibitem{ttH_1}
{Analysis of $t\bar{t}H$ and $t\bar{t}W$ production in multilepton final states
  with the ATLAS detector}.
\newblock 10 2019.

\bibitem{ttH_2}
Albert~M Sirunyan et~al.
\newblock {Measurement of the Higgs boson production rate in association with
  top quarks in final states with electrons, muons, and hadronically decaying
  tau leptons at $\sqrt{s} =$ 13 TeV}.
\newblock {\em Eur. Phys. J. C}, 81(4):378, 2021, 2011.03652.

\bibitem{tttt_1}
Georges Aad et~al.
\newblock {Evidence for $t\bar{t}t\bar{t}$ production in the multilepton final
  state in proton\textendash{}proton collisions at $\sqrt{s}=13$ $\text {TeV}$
  with the ATLAS detector}.
\newblock {\em Eur. Phys. J. C}, 80(11):1085, 2020, 2007.14858.

\bibitem{tttt_2}
Albert~M Sirunyan et~al.
\newblock {Search for production of four top quarks in final states with
  same-sign or multiple leptons in proton-proton collisions at $\sqrt{s}=$ 13
  TeV}.
\newblock {\em Eur. Phys. J. C}, 80(2):75, 2020, 1908.06463.

\bibitem{ATLAS:2021wob}
Georges Aad et~al.
\newblock {Search for new phenomena in three- or four-lepton events in $pp$
  collisions at $\sqrt{s} = 13$ TeV with the ATLAS detector}.
\newblock 7 2021, 2107.00404.

\bibitem{Sirunyan:2020tqm}
Albert~M Sirunyan et~al.
\newblock {Search for new physics in top quark production with additional
  leptons in proton-proton collisions at $\sqrt{s} = $ 13 TeV using effective
  field theory}.
\newblock {\em JHEP}, 03:095, 2021, 2012.04120.

\bibitem{EFT1}
W.~Buchmuller and D.~Wyler.
\newblock {Effective Lagrangian Analysis of New Interactions and Flavor
  Conservation}.
\newblock {\em Nucl. Phys.}, B268:621--653, 1986.

\bibitem{EFT2}
C.~Arzt, M.~B. Einhorn, and J.~Wudka.
\newblock {Patterns of deviation from the standard model}.
\newblock {\em Nucl. Phys.}, B433:41--66, 1995, hep-ph/9405214.

\bibitem{EFT3}
Martin~B. Einhorn and Jose Wudka.
\newblock {The Bases of Effective Field Theories}.
\newblock {\em Nucl. Phys.}, B876:556--574, 2013, 1307.0478.

\bibitem{EFT4}
B.~Grzadkowski, M.~Iskrzynski, M.~Misiak, and J.~Rosiek.
\newblock {Dimension-Six Terms in the Standard Model Lagrangian}.
\newblock {\em JHEP}, 10:085, 2010, 1008.4884.

\bibitem{EFT5}
Ilaria Brivio and Michael Trott.
\newblock {The Standard Model as an Effective Field Theory}.
\newblock {\em Phys. Rept.}, 793:1--98, 2019, 1706.08945.

\bibitem{U1-1}
Riccardo Barbieri, Gino Isidori, Andrea Pattori, and Fabrizio Senia.
\newblock {Anomalies in $B$-decays and $U(2)$ flavour symmetry}.
\newblock {\em Eur. Phys. J. C}, 76(2):67, 2016, 1512.01560.

\bibitem{U1-2}
Rodrigo Alonso, Benjam\'\i{}n Grinstein, and Jorge Martin~Camalich.
\newblock {Lepton universality violation and lepton flavor conservation in
  $B$-meson decays}.
\newblock {\em JHEP}, 10:184, 2015, 1505.05164.

\bibitem{U1-3}
Dario Buttazzo, Admir Greljo, Gino Isidori, and David Marzocca.
\newblock {B-physics anomalies: a guide to combined explanations}.
\newblock {\em JHEP}, 11:044, 2017, 1706.07808.

\bibitem{U1-4}
Luca Di~Luzio, Admir Greljo, and Marco Nardecchia.
\newblock {Gauge leptoquark as the origin of B-physics anomalies}.
\newblock {\em Phys. Rev. D}, 96(11):115011, 2017, 1708.08450.

\bibitem{U1-5}
Andreas Crivellin, Christoph Greub, Dario M\"uller, and Francesco Saturnino.
\newblock {Importance of Loop Effects in Explaining the Accumulated Evidence
  for New Physics in B Decays with a Vector Leptoquark}.
\newblock {\em Phys. Rev. Lett.}, 122(1):011805, 2019, 1807.02068.

\bibitem{U1-6}
Michael~J. Baker, Javier Fuentes-Mart\'\i{}n, Gino Isidori, and Matthias
  K\"onig.
\newblock {High- $p_T$ signatures in vector\textendash{}leptoquark models}.
\newblock {\em Eur. Phys. J. C}, 79(4):334, 2019, 1901.10480.

\bibitem{U1-9}
Bartosz Fornal, Sri~Aditya Gadam, and Benjamin Grinstein.
\newblock {Left-Right SU(4) Vector Leptoquark Model for Flavor Anomalies}.
\newblock {\em Phys. Rev. D}, 99(5):055025, 2019, 1812.01603.

\bibitem{U1-10}
Claudia Cornella, Javier Fuentes-Martin, and Gino Isidori.
\newblock {Revisiting the vector leptoquark explanation of the B-physics
  anomalies}.
\newblock {\em JHEP}, 07:168, 2019, 1903.11517.

\bibitem{U1-11}
P.~S. Bhupal~Dev, Rukmani Mohanta, Sudhanwa Patra, and Suchismita Sahoo.
\newblock {Unified explanation of flavor anomalies, radiative neutrino masses,
  and ANITA anomalous events in a vector leptoquark model}.
\newblock {\em Phys. Rev. D}, 102(9):095012, 2020, 2004.09464.

\bibitem{U1-12}
Syuhei Iguro, Junichiro Kawamura, Shohei Okawa, and Yuji Omura.
\newblock {TeV-scale vector leptoquark from Pati-Salam unification with
  vectorlike families}.
\newblock 3 2021, 2103.11889.

\bibitem{U1-13}
Andrei Angelescu, Damir Be\v{c}irevi\'c, Darius~A. Faroughy, Florentin
  Jaffredo, and Olcyr Sumensari.
\newblock {On the single leptoquark solutions to the $B$-physics anomalies}.
\newblock 3 2021, 2103.12504.

\bibitem{S1-S3-R2-1}
David Marzocca.
\newblock {Addressing the B-physics anomalies in a fundamental Composite Higgs
  Model}.
\newblock {\em JHEP}, 07:121, 2018, 1803.10972.

\bibitem{S1-S3-R2-2}
Hyun~Min Lee.
\newblock {Leptoquark option for B-meson anomalies and leptonic signatures}.
\newblock {\em Phys. Rev. D}, 104(1):015007, 2021, 2104.02982.

\bibitem{S1-S3-R2-3}
Chuan-Hung Chen, Takaaki Nomura, and Hiroshi Okada.
\newblock {Excesses of muon $g-2$, $R_{D^{(\ast)}}$, and $R_K$ in a leptoquark
  model}.
\newblock {\em Phys. Lett. B}, 774:456--464, 2017, 1703.03251.

\bibitem{S1-S3-R2-4}
A.~Angelescu, Damir Be\v{c}irevi\'c, D.~A. Faroughy, and O.~Sumensari.
\newblock {Closing the window on single leptoquark solutions to the $B$-physics
  anomalies}.
\newblock {\em JHEP}, 10:183, 2018, 1808.08179.

\bibitem{S1-S3-R2-5}
Innes Bigaran, John Gargalionis, and Raymond~R. Volkas.
\newblock {A near-minimal leptoquark model for reconciling flavour anomalies
  and generating radiative neutrino masses}.
\newblock {\em JHEP}, 10:106, 2019, 1906.01870.

\bibitem{S1-S3-R2-6}
Shaikh Saad.
\newblock {Combined explanations of $(g-2)_{\mu}$, $R_{D^{(*)}}$, $R_{K^{(*)}}$
  anomalies in a two-loop radiative neutrino mass model}.
\newblock {\em Phys. Rev. D}, 102(1):015019, 2020, 2005.04352.

\bibitem{S1-S3-R2-7}
K.~S. Babu, P.~S.~Bhupal Dev, Sudip Jana, and Anil Thapa.
\newblock {Unified framework for $B$-anomalies, muon $g-2$ and neutrino
  masses}.
\newblock {\em JHEP}, 03:179, 2021, 2009.01771.

\bibitem{NeubertPRL}
Martin Bauer and Matthias Neubert.
\newblock {Minimal Leptoquark Explanation for the $R_{D^{(*)}}$ , $R_K$ , and
  $(g-2)_\mu$ Anomalies}.
\newblock {\em Phys. Rev. Lett.}, 116(14):141802, 2016, 1511.01900.

\bibitem{R2_1}
Minoru Tanaka and Ryoutaro Watanabe.
\newblock {New physics in the weak interaction of $\bar B\to
  D^{(*)}\tau\bar\nu$}.
\newblock {\em Phys. Rev. D}, 87(3):034028, 2013, 1212.1878.

\bibitem{R2_2}
Ilja Dor\v{s}ner, Svjetlana Fajfer, Nejc Ko\v{s}nik, and Ivan
  Ni\v{s}and\v{z}i\'c.
\newblock {Minimally flavored colored scalar in $\bar B \to D^{(*)} \tau \bar
  \nu$ and the mass matrices constraints}.
\newblock {\em JHEP}, 11:084, 2013, 1306.6493.

\bibitem{R2_3}
Yasuhito Sakaki, Minoru Tanaka, Andrey Tayduganov, and Ryoutaro Watanabe.
\newblock {Testing leptoquark models in $\bar B \to D^{(*)} \tau \bar\nu$}.
\newblock {\em Phys. Rev. D}, 88(9):094012, 2013, 1309.0301.

\bibitem{R2_4}
Suchismita Sahoo and Rukmani Mohanta.
\newblock {Scalar leptoquarks and the rare $B$ meson decays}.
\newblock {\em Phys. Rev. D}, 91(9):094019, 2015, 1501.05193.

\bibitem{R2_5}
Chuan-Hung Chen, Takaaki Nomura, and Hiroshi Okada.
\newblock {Explanation of $B \to K^{(*)} \ell^+ \ell^-$ and muon $g-2$, and
  implications at the LHC}.
\newblock {\em Phys. Rev. D}, 94(11):115005, 2016, 1607.04857.

\bibitem{R2_6}
Ujjal~Kumar Dey, Deepak Kar, Manimala Mitra, Michael Spannowsky, and Aaron~C.
  Vincent.
\newblock {Searching for Leptoquarks at IceCube and the LHC}.
\newblock {\em Phys. Rev. D}, 98(3):035014, 2018, 1709.02009.

\bibitem{R2_7}
Damir Be\v{c}irevi\'c and Olcyr Sumensari.
\newblock {A leptoquark model to accommodate $R_K^\mathrm{exp} <
  R_K^\mathrm{SM}$ and $R_{K^\ast}^\mathrm{exp} < R_{K^\ast}^\mathrm{SM}$}.
\newblock {\em JHEP}, 08:104, 2017, 1704.05835.

\bibitem{R2_n}
Oleg Popov, Michael~A. Schmidt, and Graham White.
\newblock {$R_2$ as a single leptoquark solution to $R_{D^{(*)}}$ and
  $R_{K^{(*)}}$}.
\newblock {\em Phys. Rev. D}, 100(3):035028, 2019, 1905.06339.

\bibitem{LQ-anom1}
Pavel~Fileviez Perez, Clara Murgui, and Alexis~D. Plascencia.
\newblock {Leptoquarks and Matter Unification: Flavor Anomalies and the Muon
  $g-2$}.
\newblock 4 2021, 2104.11229.

\bibitem{Dorsner:2016wpm}
I.~Dor\v{s}ner, S.~Fajfer, A.~Greljo, J.F. Kamenik, and N.~Ko\v{s}nik.
\newblock {Physics of leptoquarks in precision experiments and at particle
  colliders}.
\newblock {\em Phys. Rept.}, 641:1--68, 2016, 1603.04993.

\bibitem{soniRPV}
Wolfgang Altmannshofer, P.~S. Bhupal~Dev, and Amarjit Soni.
\newblock {$R_{D^{(*)}}$ anomaly: A possible hint for natural supersymmetry
  with $R$-parity violation}.
\newblock {\em Phys. Rev.}, D96(9):095010, 2017, 1704.06659.

\bibitem{RPV3_1}
Christopher Brust, Andrey Katz, Scott Lawrence, and Raman Sundrum.
\newblock {SUSY, the Third Generation and the LHC}.
\newblock {\em JHEP}, 03:103, 2012, 1110.6670.

\bibitem{RPV3_2}
Michele Papucci, Joshua~T. Ruderman, and Andreas Weiler.
\newblock {Natural SUSY Endures}.
\newblock {\em JHEP}, 09:035, 2012, 1110.6926.

\bibitem{soniRPV2}
Wolfgang Altmannshofer, P.~S.~Bhupal Dev, Amarjit Soni, and Yicong Sui.
\newblock {Addressing R$_{D^{(*)}}$, R$_{K^{(*)}}$, muon $g-2$ and ANITA
  anomalies in a minimal $R$-parity violating supersymmetric framework}.
\newblock {\em Phys. Rev. D}, 102(1):015031, 2020, 2002.12910.

\bibitem{soniRPV3}
P.~S. Bhupal~Dev, Amarjit Soni, and Fang Xu.
\newblock {Hints of Natural Supersymmetry in Flavor Anomalies?}
\newblock 6 2021, 2106.15647.

\bibitem{StraubEFT}
Martin Jung and David~M. Straub.
\newblock {Constraining new physics in $b\to c\ell\nu$ transitions}.
\newblock {\em JHEP}, 01:009, 2019, 1801.01112.

\bibitem{madgraph5}
Johan Alwall, Michel Herquet, Fabio Maltoni, Olivier Mattelaer, and Tim
  Stelzer.
\newblock {MadGraph 5 : Going Beyond}.
\newblock {\em JHEP}, 06:128, 2011, 1106.0522.

\bibitem{ATLAS:2021eyc}
{Search for new phenomena in three- or four-lepton events in $pp$ collisions at
  $\sqrt{s} = $ 13 TeV with the ATLAS detector}.
\newblock 3 2021, ATLAS-CONF-2021-011.

\bibitem{FRpaper}
Adam Alloul, Neil~D. Christensen, Céline Degrande, Claude Duhr, and Benjamin
  Fuks.
\newblock {FeynRules 2.0 - A complete toolbox for tree-level phenomenology}.
\newblock {\em Comput. Phys. Commun.}, 185:2250--2300, 2014, 1310.1921.

\bibitem{Ball:2014uwa}
Richard~D. Ball et~al.
\newblock {Parton distributions for the LHC Run II}.
\newblock {\em JHEP}, 04:040, 2015, 1410.8849.

\bibitem{Mrenna:2016sih}
S.~Mrenna and P.~Skands.
\newblock {Automated Parton-Shower Variations in Pythia 8}.
\newblock {\em Phys. Rev. D}, 94(7):074005, 2016, 1605.08352.

\bibitem{MLM}
Michelangelo~L. Mangano, Mauro Moretti, Fulvio Piccinini, and Michele Treccani.
\newblock {Matching matrix elements and shower evolution for top-quark
  production in hadronic collisions}.
\newblock {\em JHEP}, 01:013, 2007, hep-ph/0611129.

\bibitem{deFavereau:2013fsa}
J.~de~Favereau, C.~Delaere, P.~Demin, A.~Giammanco, V.~Lemaître, A.~Mertens,
  and M.~Selvaggi.
\newblock {DELPHES 3, A modular framework for fast simulation of a generic
  collider experiment}.
\newblock {\em JHEP}, 02:057, 2014, 1307.6346.

\bibitem{Cacciari:2008gp}
Matteo Cacciari, Gavin~P. Salam, and Gregory Soyez.
\newblock {The anti-$k_t$ jet clustering algorithm}.
\newblock {\em JHEP}, 04:063, 2008, 0802.1189.

\bibitem{Cacciari:2011ma}
Matteo Cacciari, Gavin~P. Salam, and Gregory Soyez.
\newblock {FastJet User Manual}.
\newblock {\em Eur. Phys. J.}, C72:1896, 2012, 1111.6097.

\bibitem{Cacciari:2005hq}
Matteo Cacciari and Gavin~P. Salam.
\newblock {Dispelling the $N^{3}$ myth for the $k_t$ jet-finder}.
\newblock {\em Phys. Lett.}, B641:57--61, 2006, hep-ph/0512210.

\bibitem{ATL-PHYS-PUB-2015-022}
{Expected performance of the ATLAS $b$-tagging algorithms in Run-2}.
\newblock Technical Report ATL-PHYS-PUB-2015-022, CERN, Geneva, Jul 2015.

\bibitem{Verkerke:2003ir}
Wouter Verkerke and David~P. Kirkby.
\newblock {The RooFit toolkit for data modeling}.
\newblock {\em eConf}, C0303241:MOLT007, 2003, physics/0306116.
\newblock [,186(2003)].

\bibitem{Read:2002hq}
Alexander~L. Read.
\newblock {Presentation of search results: The CL(s) technique}.
\newblock {\em J. Phys.}, G28:2693--2704, 2002.
\newblock [,11(2002)].

\bibitem{Dawson:2018dxp}
S.~Dawson, P.P. Giardino, and A.~Ismail.
\newblock {Standard model EFT and the Drell-Yan process at high energy}.
\newblock {\em Phys. Rev. D}, 99(3):035044, 2019, 1811.12260.

\end{thebibliography}

\newpage

\appendix

\section{Cut-flow tables}

We present below cut-flow tables of the number of events for the signals and the dominant 
backgrounds considered in this work, in the di-, tri- and four-lepton channels, in Tables~\ref{tab:cutflow_2l},~\ref{tab:cutflow_3l},~\ref{tab:cutflow_4l}.

In all cases we assume an integrated luminosity of ${\cal L}=300$~fb$^{-1}$ and start with a baseline value of 
$m_{\mu^+ \mu^-}^{\tt min} = 300$ GeV for the di-muon invariant mass lower cut. Then, for each case we list the number of events after imposing the optimal di-muon invariant mass lower cut and the chosen jet selections (see main text for further details).

\begin{widetext}

\begin{table}[htb]
\centering
\begin{tabular}{ |l|c|c|c|c|c| }
\hline
\multirow{1}{*}{Cuts} &
      \multicolumn{2}{|c|}{Background Events} &
      \multicolumn{3}{|c|}{Signal Events} \\
\hline
Process & $t \bar t Z$ (irreducible) & ~~~~~~~$t \bar t$~~~~~~~ & ~~~~~$T_{RR} = 1$~~~~~ & ~~$V_{RR} = 1~(-1)$~~ & ~~~~~$S_{RR} = 1$~~~~~ \\
\hline
$m_{\mu^+ \mu^-}^{\tt min} = 300$~GeV & 9.1 & 6221.4 & 379.0 & 44.7 (33.4) & 9.2 \\
\hline
$m_{\mu^+ \mu^-}^{\tt min} = 1400$~GeV & 0.0 & 1.1 & 36.8 & 4.4 (4.2) & 2.0 \\
\hline
N$_{j} \geq$ 3, N$_{b} \geq 1$ & 0.0 & 0.5 & 22.7 & 3.4 (3.3) & 1.4 \\
\hline

\end{tabular}
\caption{Cut-flow table for the selection of 2 leptons. The number of events appears after each selection is for an integrated luminosity of ${\cal L}=300$~fb$^{-1}$.}
\label{tab:cutflow_2l}
\end{table}

\begin{table}[htb]
\centering
\begin{tabular}{ |l|c|c|c|c|c| }
\hline
\multirow{1}{*}{Cuts} &
      \multicolumn{2}{|c|}{Background Events} &
      \multicolumn{3}{|c|}{Signal Events} \\
\hline
Process & $t \bar t Z$ (irreducible) & ~~~~~$WZ$~~~~~ & ~~~~~$T_{RR} = 1$~~~~~ & ~~$V_{RR} = 1~(-1)$~~ & ~~~~~$S_{RR} = 1$~~~~~ \\
\hline
$m_{\mu^+ \mu^-}^{\tt min} = 300$~GeV & 4.8 & 45.5  & 91.4 & 13.5 (10.2) & 2.1 \\
\hline
$m_{\mu^+ \mu^-}^{\tt min} = 500$~GeV & 0.8 & 12.5 & 52.4 & 8.2 (7.1) & 1.5 \\
\hline
N$_{j} \geq$ 2, N$_{b} \geq 1$ & 0.4 & 0.2 & 35.8 & 6.1 (5.4) & 1.1 \\
\hline
\end{tabular}
\caption{Cut-flow table for the selection of 3 leptons. The number of events appears after each selection is for an integrated luminosity of ${\cal L}=300$~fb$^{-1}$.}
\label{tab:cutflow_3l}
\end{table}

\begin{table}[htb]
\centering
\begin{tabular}{ |l|c|c|c|c|c| }
\hline
\multirow{1}{*}{Cuts} &
      \multicolumn{2}{|c|}{Background Events} &
      \multicolumn{3}{|c|}{Signal Events} \\
\hline
Process & $t \bar t Z$ (irreducible) & ~~~~~$ZZ$~~~~~ & ~~~~~$T_{RR} = 1$~~~~~ & ~~$V_{RR} = 1~(-1)$~~ & ~~~~~$S_{RR} = 1$~~~~~ \\
\hline
$m_{\mu^+ \mu^-}^{\tt min} = 300$~GeV & 0.8 & 3.5 & 5.7 & 1.2 (0.9) & 0.2 \\
\hline
N$_{j} \geq$ 2, N$_{b} \geq 1$ & 0.0 & 0.0 & 3.6 & 0.8 (0.6) & 0.1 \\
\hline
\end{tabular}
\caption{Cut-flow table for the selection of 4 leptons. The number of events appears after each selection is for an integrated luminosity of ${\cal L}=300$~fb$^{-1}$.}
\label{tab:cutflow_4l}
\end{table}

\end{widetext}

\end{document}